\documentclass[11pt,onecolumn,journal]{IEEEtran}
\usepackage{amsmath,amsfonts}
\usepackage{algorithmic}
\usepackage{algorithm}
\usepackage{array}
\usepackage[caption=false,font=normalsize,labelfont=sf,textfont=sf]{subfig}
\usepackage{textcomp}
\usepackage{stfloats}
\usepackage{url}
\usepackage{verbatim}
\usepackage{graphicx}
\usepackage{glossaries}
\usepackage{cite}
\usepackage{hyperref}
\usepackage{booktabs}
\usepackage{subfig}
\usepackage{float}
\usepackage{tikz}
\usepackage{xspace}

\usepackage{booktabs}
\usepackage{tabularx}

\newacronym{anm}{ANM}{Additive Noise Model}
\newacronym{ccm}{CCM}{Convergent Cross-Mapping}
\newacronym{ci}{CI}{Conditional Independence}
\newacronym{cte}{CTE}{Conditional Transfer Entropy}
\newacronym{dag}{DAG}{Directed Acyclic Graph}
\newacronym{dgp}{DGP}{Data Generating Process}
\newacronym{gc}{GC}{Granger Causality}
\newacronym{ica}{ICA}{Independent Component Analysis}
\newacronym{icu}{ICU}{Intensive-Care Unit}
\newacronym{mci}{MCI}{Momentary Conditional Independence}
\newacronym{ml}{ML}{Machine Learning}
\newacronym{ode}{ODE}{ordinary differential equation}
\newacronym{sem}{SEM}{Structural Equation Model}
\newacronym{shd}{SHD}{Structural Hamming Distance}
\newacronym{sid}{SID}{Structural Intervention Distance}
\newacronym{surd}{SURD}{Synergistic-Unique-Redundant Decomposition of causality}
\newacronym{te}{TE}{Transfer Entropy}
\newacronym{tscd}{TSCD}{Time-Series Causal Discovery}
\newacronym{var}{VAR}{Vector Autoregressive}

\newcommand{\ie}{i.e.\@\xspace}
\newcommand{\indep}{\perp \!\!\! \perp}

\begin{document}

% tikzlibrary.code.tex
%
% Copyright 2010-2011 by Laura Dietz
% Copyright 2012 by Jaakko Luttinen
%
% The MIT License
%
% See LICENSE file for more details.
%
% This has been modified by Michael Camilleri

% Load other libraries
\usetikzlibrary{shapes}
\usetikzlibrary{fit}
\usetikzlibrary{chains}
\usetikzlibrary{arrows}

% Latent node
\tikzstyle{latent} = [circle,fill=white,draw=black,inner sep=2pt,
minimum size=30pt, font=\fontsize{10}{10}\selectfont, node distance=1]
\tikzstyle{latent_big} = [latent, minimum size=40pt]
\tikzstyle{latent_med} = [latent, minimum size=35pt]
\tikzstyle{latent_sml} = [latent, minimum size=20pt, font=\fontsize{8}{8}\selectfont]
% Observed node
\tikzstyle{obs} = [latent,fill=gray!25]
\tikzstyle{obs_big} = [obs, minimum size=45pt]
\tikzstyle{obs_med} = [obs, minimum size=35pt]
\tikzstyle{obs_sml} = [obs, minimum size=25pt, font=\fontsize{8}{8}\selectfont]
% Constant node
\tikzstyle{const_lat} = [rectangle, draw=black, inner sep=2pt, node distance=1, fill=white, minimum size=15pt]
\tikzstyle{const} = [rectangle, draw=black, inner sep=2pt, node distance=1, fill=gray!25, minimum size=15pt]
% Blank: for adding dots
\tikzstyle{blank} = [rectangle, inner sep=0pt, node distance=1]
% Factor node
\tikzstyle{factor} = [rectangle, fill=black, minimum size=5pt, inner
sep=0pt, node distance=0.4]
% Deterministic node
\tikzstyle{det} = [latent, diamond]

% Plate node
\tikzstyle{plate} = [draw, rectangle, rounded corners, fit=#1, inner sep=5pt]
% Invisible wrapper node
\tikzstyle{wrap} = [inner sep=0pt, fit=#1]
% Gate
\tikzstyle{gate} = [draw, rectangle, dashed, fit=#1]

% Caption node
\tikzstyle{caption} = [font=\footnotesize, node distance=0] %
\tikzstyle{plate caption} = [caption, node distance=0, inner sep=0pt,
right=5pt of #1.south east] %
\tikzstyle{factor caption} = [caption] %
\tikzstyle{every label} += [caption] %

%\pgfdeclarelayer{b}
%\pgfdeclarelayer{f}
%\pgfsetlayers{b,main,f}

% \factoredge [options] {inputs} {factors} {outputs}
\newcommand{\factoredge}[4][]{ %
  % Connect all nodes #2 to all nodes #4 via all factors #3.
  \foreach \f in {#3} { %
    \foreach \x in {#2} { %
      \path (\x) edge[-,#1] (\f) ; %
      %\draw[-,#1] (\x) edge[-] (\f) ; %
    } ;
    \foreach \y in {#4} { %
      \path (\f) edge[->, >={triangle 45}, #1] (\y) ; %
      %\draw[->,#1] (\f) -- (\y) ; %
    } ;
  } ;
}

% \edge [options] {inputs} {outputs}
\newcommand{\edge}[3][]{ %
  % Connect all nodes #2 to all nodes #3.
  \foreach \x in {#2} { %
    \foreach \y in {#3} { %
      \path (\x) edge [->, >={triangle 45}, #1] (\y) ;%
      %\draw[->,#1] (\x) -- (\y) ;%
    } ;
  } ;
}

\newcommand{\dottededge}[3][]{ %
  % Connect all nodes #2 to all nodes #3.
  \foreach \x in {#2} { %
    \foreach \y in {#3} { %
      \path (\x) edge [dashed, <->, >={triangle 45}, #1] (\y) ;%
      %\draw[->,#1] (\x) -- (\y) ;%
    } ;
  } ;
}

% \factor [options] {name} {caption} {inputs} {outputs}
\newcommand{\factor}[5][]{ %
  % Draw the factor node. Use alias to allow empty names.
  \node[factor, label={[name=#2-caption]#3}, name=#2, #1,
  alias=#2-alias] {} ; %
  % Connect all inputs to outputs via this factor
  \factoredge {#4} {#2-alias} {#5} ; %
}

% \plate [options] {name} {fitlist} {caption}
\newcommand{\plate}[4][]{ %
  \node[wrap=#3] (#2-wrap) {}; %
  \node[plate caption=#2-wrap] (#2-caption) {#4}; %
  \node[plate=(#2-wrap)(#2-caption), #1] (#2) {}; %
}

% \gate [options] {name} {fitlist} {inputs}
\newcommand{\gate}[4][]{ %
  \node[gate=#3, name=#2, #1, alias=#2-alias] {}; %
  \foreach \x in {#4} { %
    \draw [-*,thick] (\x) -- (#2-alias); %
  } ;%
}

% \vgate {name} {fitlist-left} {caption-left} {fitlist-right}
% {caption-right} {inputs}
\newcommand{\vgate}[6]{ %
  % Wrap the left and right parts
  \node[wrap=#2] (#1-left) {}; %
  \node[wrap=#4] (#1-right) {}; %
  % Draw the gate
  \node[gate=(#1-left)(#1-right)] (#1) {}; %
  % Add captions
  \node[caption, below left=of #1.north ] (#1-left-caption)
  {#3}; %
  \node[caption, below right=of #1.north ] (#1-right-caption)
  {#5}; %
  % Draw middle separation
  \draw [-, dashed] (#1.north) -- (#1.south); %
  % Draw inputs
  \foreach \x in {#6} { %
    \draw [-*,thick] (\x) -- (#1); %
  } ;%
}

% \hgate {name} {fitlist-top} {caption-top} {fitlist-bottom}
% {caption-bottom} {inputs}
\newcommand{\hgate}[6]{ %
  % Wrap the left and right parts
  \node[wrap=#2] (#1-top) {}; %
  \node[wrap=#4] (#1-bottom) {}; %
  % Draw the gate
  \node[gate=(#1-top)(#1-bottom)] (#1) {}; %
  % Add captions
  \node[caption, above right=of #1.west ] (#1-top-caption)
  {#3}; %
  \node[caption, below right=of #1.west ] (#1-bottom-caption)
  {#5}; %
  % Draw middle separation
  \draw [-, dashed] (#1.west) -- (#1.east); %
  % Draw inputs
  \foreach \x in {#6} { %
    \draw [-*,thick] (\x) -- (#1); %
  } ;%
} % Allows drawing DAGs

\title{One Thing Leads to Another: The Essentials of Time Series Causal Discovery}

\author{Kurt Butler,~\IEEEmembership{Member,~IEEE},
Damian Machlanski,
Michael P.J. Camilleri,
Zhihua Liu,
Panagiotis Dimitrakopoulos,
Feng Chen,
Ricardo Silva, and
Sotirios A. Tsaftaris, ~\IEEEmembership{Senior~Member,~IEEE}
% $^1$School of Engineering, The University of Edinburgh \\
% $^2$Causality in Healthcare AI Hub (CHAI) \\
% $^3$Department of Statistical Science, University College London
        % <-this % stops a space
% \thanks{This document is prepared by the Special Issue Area Editor, Selin Aviyente  and by the Editor-in-Chief, T\"{u}lay~Adal{\i}.}% <-this % stops a space
\thanks{Kurt Butler, Damian Machlanski, Zhihua Liu, Panagiotis Dimitrakopoulos and Sotirios A. Tsaftaris are with the School of Engineering, The University of Edinburgh, and the Causality in Healthcare AI Hub (CHAI). Michael P.J. Camilleri is with the School of Informatics, The University of Edinburgh and CHAI. Feng Chen is with the School of Engineering, The University of Edinburgh. Ricardo Silva is with the Department of Statistical Science, University College London, and CHAI.}% <-this % stops a space
% \thanks{Special Issue Area Editor email: SPM-SI-AREA@LISTSERV.IEEE.ORG.}
% \thanks{Manuscript received July 1, 2026.}
\thanks{This manuscript has been accepted for
publication in IEEE Signal Processing Magazine.}
\thanks{\copyright\ 2026 IEEE. Personal use of this material is permitted.  Permission from IEEE must be obtained for all other uses, in any current or future media, including reprinting/republishing this material for advertising or promotional purposes, creating new collective works, for resale or redistribution to servers or lists, or reuse of any copyrighted component of this work in other works.}}

\markboth{IEEE Signal Processing Magazine,~Vol.~XX, No.~XX, June~2026}%
{Aviyente \MakeLowercase{\textit{et al.}}: Author Guidelines for Special Issue Articles of IEEE SPM}

\maketitle

\begin{abstract}
    Time series causal discovery --- the process of detecting causation between time series --- has its roots in signal processing research from over a half century ago. 
    It has since received active and widespread interest, evolving into different forms across domains such as statistics, dynamical systems and machine learning. Entering this fragmented and constantly evolving research landscape can be daunting.
    This tutorial provides a unified overview of time series causal discovery across research communities. We introduce the key principles invoked by modern discovery methods, and we discuss their limitations and assumptions. We also discuss how discovery methods are evaluated, benchmarked and fine-tuned. We further discuss how causal discovery for time series is becoming crucial for downstream tasks in agriculture and Earth science, in healthcare, and in artificial intelligence systems. We believe that this tutorial will serve as a valuable resource in navigating the deluge of recent causality research, as well as an invitation to signal processing audiences to contribute to causal problems. To further help readers, we host a companion repository to link to useful code packages and resources at \url{https://github.com/CHAI-UK/Awesome-Time-Series-Causal-Discovery}.
\end{abstract}

\section{Introduction}
%  will motivate the problem of time series causal discovery and it's relevance to science and engineering. 
% We call attention to the wide variety of approaches to the problem \cite{granger1969investigating,pearl2009causality, runge2019inferring, assaad2022survey} and observe that these methods often make conflicting assumptions with unclear guidance on best practices for causal discovery. We resolve these uncertainties by advocating for a temporal perspective on causality to frame the problem.

The modern study of causality began with time series. Problems of causality have been discussed for thousands of years, as early as Aristotle, and later Hume. It was however the work of Granger \cite{granger1969investigating} which popularized the problem of detecting causation between time series, which we call \textit{causal discovery}. Granger's theory has since found widespread interest and applications across the sciences \cite{bressler2011wiener}. 
%Across academic communities, the task of ``discovering causal relationships'' has taken on different meanings, and many communities appear to have different standards and protocols for concluding causation.

Granger's work depended upon the idea that a cause always precedes its effect, a principle sometimes called \textit{temporal precedence} or the ``arrow of time''. This approach was naturally tailored to the time series setting, and as a result Granger causality has inspired many approaches to time series causal discovery.
% However, causality is a topic of interest in 
% The problem of causal discovery is attacked differently for static or temporal variables. In the static case, the goal is to detect a causal influence between two or more random variables. 
However, time series are a unique modality: temporal information is embedded in the structure of the data. For other data modalities such as tabular data or images, Granger's principle is not directly applicable. Thus, to apply causal reasoning to other data types, other schools of thought emerged. 

Most modern definitions of causality disagree with Granger since his notion of causality is not defined by how a system behaves under intervention. An \textbf{intervention} is a process in which we modify the \gls{dgp} of a system, say, by treating a sick patient or controlling the behaviour of a specific variable. The most common definition these days comes from Pearl \cite{pearl2009causality}, where causation means that intervention upon the cause variable should affect the affected variable. These interventions need only be possible in principle, not necessarily in real life, to establish a causal link. Causality simply means that if you could change the cause, the effect would change too.
Interventions in this framework are formalized mathematically using the $do$-operator (Section \ref{sec:pearl}). 

For Pearl \cite{pearl2009causality} and others, it is natural to represent causal relationships using a directed graph. The discovery problem is then framed as a graph topology inference task, where the nodes of the graph are random variables, and edges represent causal dependencies. Importantly, these causal graphs are acyclic. Causality is inherently temporal, and physics tells us that causal influences only propagate forward in time. Cycles in the graph would then violate this principle. Pearl's approach has also become widely adopted, and the \textbf{\gls{dag}} principle has become the \textit{de facto} approach to causality in the machine learning literature. Eventually, as the Pearlian approach was extended and developed, extensions of DAG-based methods were proposed for time series. 

Today there are many methods for \gls{tscd}. Many of these methods were developed using distinct schools of thought, each using their own principles and definitions of causality, although they all attempt to solve the same problem. It can be very unclear in practice how to get started with \gls{tscd}, and how to pick a method to use for a given problem. 
In this tutorial, we present a comprehensive introduction to \gls{tscd}. Readers new to causality can expect to leave this tutorial able to navigate the literature, to recognize which principles underlie a given method, and to critically evaluate \gls{tscd} methods for their own problems. Readers seeking implementation-level details are recommended to pair this tutorial with the companion repository and the primary references that we cite throughout.

\subsection{Limitations and scope}
A full tutorial and review of \gls{tscd} would require an entire book. This tutorial is necessarily selective, covering only a few representative principles and methods within each category. Many capable algorithms, extensions, and alternative principles have been excluded. We also do not comment on specialized methods that address real-world challenges such as missing data, irregular sampling, or non-stationarity. Additional \gls{tscd} methods, software packages and related papers are linked in the companion repository to this paper, which we will continue to update as the field evolves. 
For a deeper, more technical survey of \gls{tscd} methods, we recommend complementing this tutorial with more in-depth reviews including \cite{assaad2022survey}.

\subsection{How do I get started with \gls{tscd}?}
For readers looking for an immediate programming-oriented entry to \gls{tscd}, we provide an online companion quick-start guide, accessible via the link provided in the abstract.

The remainder of this article explores the theory, methods, and broader impacts of \gls{tscd}, organized as follows. 
In Section \ref{sec:preliminaries}, we introduce the basic concepts, notation and problem formulation required to understand the \gls{tscd} task.
In Section \ref{sec:methods}, we compare and contrast principles for \gls{tscd}, and we discuss some representative algorithms that operationalize these principles. 
In Section \ref{sec:Best Practices for Evaluation and Tuning}, we discuss some best practices for hyperparameter tuning, evaluation metrics and other considerations.
In Section \ref{sec:impact}, we discuss how causal discovery, as a detection task, plays a critical role in high-stakes and emerging applications such as healthcare and AI.
In Section \ref{sec:outlook}, we comment on remaining challenges and recent developments in causality research, and the role signal processing has to play in these fields.
Section \ref{sec:conclusion} concludes the paper.

\section{Preliminaries and Problem Statement}
\label{sec:preliminaries}
We start by defining the mathematical notation and formulating the space of problems we will consider.
% To make this more concrete we will motivate it by the running example of a patient in an \gls{icu} ward, monitored through multiple sensors.

\subsection{Basic notation}
% We begin by defining the notion of a (possibly stochastic) variable $X$ which represents a quantity we wish to model.
% In our example, $X$ could be a heartbeat monitor or a blood oxygen level sensor.
% $X$ is characterised by:
% \begin{itemize}
%     \item its support: \ie the range of values it can take. We distinguish primarily between discrete (categorical) variables taking values in some finite set and continuous variables defined over the space of real-values (which itself may be bounded or unbounded).
%     \item the dimensionality: \ie $X$ can be a single scalar value or a multi-variate variable which is modelled jointly.
% \end{itemize}
In this tutorial, we will be primarily interested in (possibly stochastic) time series $X_t$ indexed by a discrete time index $t=1,...,T$. Depending on the context, we adopt one of two conventions for describing signals.
In the case that we have only two or three time series of interest, we may denote them as $X_t$, $Y_t$ and $Z_t$.
For more variables, we will describe them compactly using $X_{i,t}$, where $i=1,...,d$ enumerates each signal: in the case that we need to refer to them collectively, we use bold notation, $\mathbf{X}_t=[X_{1,t},X_{2,t},\cdots,X_{d,t}]$.
Unless otherwise stated, it is assumed that $X_{i,t} \in \mathbb{R}$.
% In our setup, we consider a number of these variables $X_{i,t}$, where $i \in \left\lbrace1, \ldots, I\right\rbrace$ is an index over different variables and $t \in \left\lbrace1,\ldots,T\right\rbrace$ is an index over time.
% In our patient example, this would constitute all the sensors for the subject, and we refer to this system (subject) observed over time as a time-series.
% Note that the time index of temporal variables can also be continuous in time (even if observed only at a finite sample of time points), in which case these are modelled as dynamical systems, but these are outside the scope of our review.

We assume that we only have one realization of a system to study, meaning one set of time series. Most methods for \gls{tscd} are designed for such cases. Methods that exploit multiple realizations are beyond the scope of this tutorial.
% However, there may be other patients in the same ward, and it may be reasonable to assume that each subject (system), although independent, follows the same biological process (human physiology), and hence we can learn about this system by observing all patients.
% This adds another layer of indexing to our system $X^{[n]}_{i,t}$ where $n \in \left\lbrace1,\ldots,N\right\rbrace$ is an index over the independent subjects.
% However, we will mostly consider methods which target the single-realisation, because they are the most common case encountered in the literature.

% \begin{table}[t]
% \centering
% \caption{List of Acronyms}
% \label{tab:acronyms}
% \begin{tabular}{ll}
% \toprule
% \textbf{Acronym} & \textbf{Definition} \\
% \midrule
% \gls{anm}       & \glsentrylong{anm} \\
% \gls{ccm}       & \glsentrylong{ccm} \\
% \gls{ci}        & \glsentrylong{ci} \\
% \gls{dag}       & \glsentrylong{dag} \\
% \gls{dgp}       & \glsentrylong{dgp} \\
% \gls{gc}        & \glsentrylong{gc} \\
% \gls{ica}       & \glsentrylong{ica} \\
% \gls{mci}       & \glsentrylong{mci} \\
% \gls{shd}       & \glsentrylong{shd} \\
% \gls{sid}       & \glsentrylong{sid} \\
% \gls{tscd}      & \glsentrylong{tscd} \\
% \bottomrule
% \end{tabular}
% \end{table}

\subsection{Graphs and graph notation}
\begin{figure}
    \centering
    \subfloat[Full Graph]{\begin{tikzpicture}
    % =============== Variables (Nodes) =============== %
    \node[obs_sml]                      (X11)   {$X_{1,1}$};
    \node[obs_sml, below=0.7cm of X11]  (X21)   {$X_{2,1}$};
    \node[obs_sml, above=0.7cm of X11]  (X41)   {$X_{4,1}$};

    \node[obs_sml, right=1.5cm of X11] (X12)   {$X_{1,2}$};
    \node[obs_sml, below=0.7cm of X12] (X22)   {$X_{2,2}$};
    \node[obs_sml, below=0.7cm of X22] (X32)   {$X_{3,2}$};
    \node[obs_sml, above=0.7cm of X12] (X42)   {$X_{4,2}$};

    \node[obs_sml, right=1.5cm of X12] (X13)   {$X_{1,3}$};
    \node[obs_sml, below=0.7cm of X13] (X23)   {$X_{2,3}$};
    \node[obs_sml, below=0.7cm of X23] (X33)   {$X_{3,3}$};
    \node[obs_sml, above=0.7cm of X13] (X43)   {$X_{4,3}$};

    \node[obs_sml, right=1.5cm of X13] (X14)   {$X_{1,4}$};
    \node[obs_sml, below=0.7cm of X14] (X24)   {$X_{2,4}$};
    \node[obs_sml, below=0.7cm of X24] (X34)   {$X_{3,4}$};

    % =============== Causality (Edges) =============== %
    \edge {X11} {X12}
    \edge {X12} {X13}
    \edge {X13} {X14}

    \edge {X11} {X21}
    \edge {X11} {X32}

    \edge {X12} {X22}
    \edge {X12} {X33}

    \edge {X13} {X23}
    \edge {X13} {X34}

    \edge {X14} {X24}

    \edge {X41} {X12}
    \edge {X41} {X42}
    
    \edge {X42} {X13}
    \edge {X42} {X43}
    
    \edge {X43} {X14}

\end{tikzpicture}}
    \hfill
    \subfloat[Summary Graph]{\begin{tikzpicture}
    \node[obs_sml]                     (X1t)   {$X_{1,t}$};
    \node[obs_sml, below=0.7cm of X1t] (X2t)   {$X_{2,t}$};
    \node[obs_sml, below=0.7cm of X2t] (X3t)   {$X_{3,t}$};
    \node[obs_sml, above=0.7cm of X1t] (X4t)   {$X_{4,t}$};
    \node[blank, left=1.5cm of X1t]      (p)     {};
    \node[blank, right=1.5cm of X1t]     (n)     {};
    
    \edge [loop left]  {X4t} {X4t}
    \edge              {X4t} {X1t}
    \edge [loop left]  {X1t} {X1t}
    \edge              {X1t} {X2t}
    \edge [bend right] {X1t} {X3t}
\end{tikzpicture}}
    \hfill
    \subfloat[Window Graph]{\begin{tikzpicture}
    % =============== Variables (Nodes) =============== %
    \node[obs_sml]                     (X1p)   {$X_{1,t'}$};
    \node[obs_sml, below=0.7cm of X1p] (X2p)   {$X_{2,t'}$};
    \node[obs_sml, above=0.7cm of X1p] (X4p)   {$X_{4,t'}$};

    \node[obs_sml, right=1.5cm of X1p] (X1t)   {$X_{1,t}$};
    \node[obs_sml, below=0.7cm of X1t] (X2t)   {$X_{2,t}$};
    \node[obs_sml, below=0.7cm of X2t] (X3t)   {$X_{3,t}$};
    \node[obs_sml, above=0.7cm of X1t] (X4t)   {$X_{4,t}$};

    % =============== Causality (Edges) =============== %
    \edge {X1p} {X1t}

    \edge {X1p} {X2p}
    \edge {X1p} {X3t}

    \edge {X1t} {X2t}

    \edge {X4p} {X1t}
    \edge {X4p} {X4t}
    
\end{tikzpicture}}
    \caption{Three ways to graphically represent causality between time series. Unlike Full Graphs and Window Graphs, Summary Graphs are permitted to have cycles.}
    \label{fig:example_dag}
\end{figure}
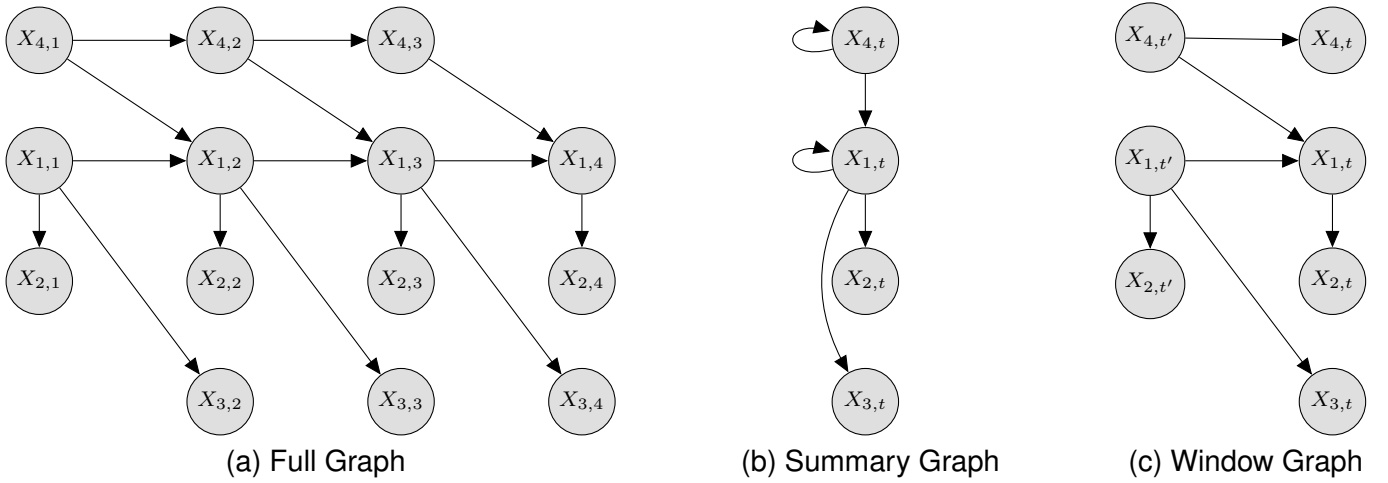

It is often helpful to represent a causal system by a \gls{dag}, in which different variables in $\textbf{X}$ are represented by the nodes, and the directed edges indicate the direction of the causal relationship \cite{pearl2009causality}.
An example \gls{dag} for a hypothetical patient is shown in Fig.\ \ref{fig:example_dag}.
In this case $X_{1,t}$ is the patient state, $X_{2,t}$, $X_{3,t}$ are sensors, and $X_{4,t}$ is the flow of an intravenous drug.
There are multiple ways in which we can represent this setup.
In Fig. \ref{fig:example_dag}(a), the \textbf{full graph} shows the state of all variables over the entire observation period.
It is the most flexible, but also intractable for any but the shortest of time-series.
Alternatively, the \textbf{summary graph} in Fig. \ref{fig:example_dag}(b) is a concise representation which shows only which variables cause other variables but there is no notion of time lag (note for example how $X_4$ causes $X_1$ with a delay of 1 time-step).
In between these representations, the \textbf{window graph} of Fig. \ref{fig:example_dag}(c) shows the minimal set of time-points to represent the causal effects and time-delays at which they act.
In particular, if the process is stationary, the window graph suffices to fully characterise the entire process. Furthermore, note that while full and window graphs are DAGs, summary graphs can have cycles.
Unless otherwise stated, we use the window graph as our default representation in this tutorial.

In text, we write $X_{i,s} \to X_{j,t}$ to refer to the corresponding edge in the causal graph. Given a graph $G$, the \textbf{parent set} $Pa(X_{j,t})$ is the set of all nodes $X_{i,s}$ such that $X_{i,s} \to X_{j,t}$ exists in $G$. 

\subsection{Interventions and structural causality}
\label{sec:pearl}
Pearl's theory of causality frames causal models as DGPs that can be affected by interventions \cite{pearl2009causality}. DGPs can be represented either as a system of structural equations (like a generative algorithm), or as a factorization of the joint probability distribution \cite[\S 3.1]{peters2017elements}. In both cases, the order in which variables are generated must be consistent with the partial ordering induced by the full graph. The do-operator is a formal operator which represents certain interventions (changes to the DGP) \cite[\S 3.2]{peters2017elements}. The notation $\mathrm{do}(X_t=x_t)$ denotes the intervention in which, at time $t$, the value of $X_t$ is clamped to exactly $x_t$. The intervention $\mathrm{do}(X_t=x_t)$ also affects the (full time) causal graph by deleting any edges that enter $X_t$. Generally, $p(Y_t|\mathrm{do}(X_t=x_t))$ and $p(Y_t|X_t=x_t)$ are different probability distributions, since they correspond to different DGPs. However, a common trick to estimate $p(Y_t|\mathrm{do}(X_t=x_t))$ from it's non-interventional counterpart is to use an \textbf{adjustment set}, a set of variables $\mathbf{Z}_t$ which has the property that
\begin{equation}
p(Y_t|\mathrm{do}(X_t=x_t),\mathbf{Z}_t = \mathbf{z}_t)
=
p(Y_t|X_t=x_t,\mathbf{Z}_t = \mathbf{z}_t).
\end{equation}
The question of whether $\mathbf{Z}_t$ is a valid adjustment set can be answered solely using the topology of the window graph \cite[pp.107-118]{ peters2017elements}. It is possible that $\mathbf{Z}_t = \emptyset$, which happens (for example) when $Pa(X_t) = \emptyset$.

We will not touch on adjustment sets further in this tutorial, but we mention them to motivate how methods defined in terms of interventional causality are implemented without access to interventional data. This logic implicitly justifies many of the methods shown in Table \ref{tab:methods}.

\subsection{Problem statement}
\begin{figure}
    \centering
    \includegraphics[width=0.95\linewidth]{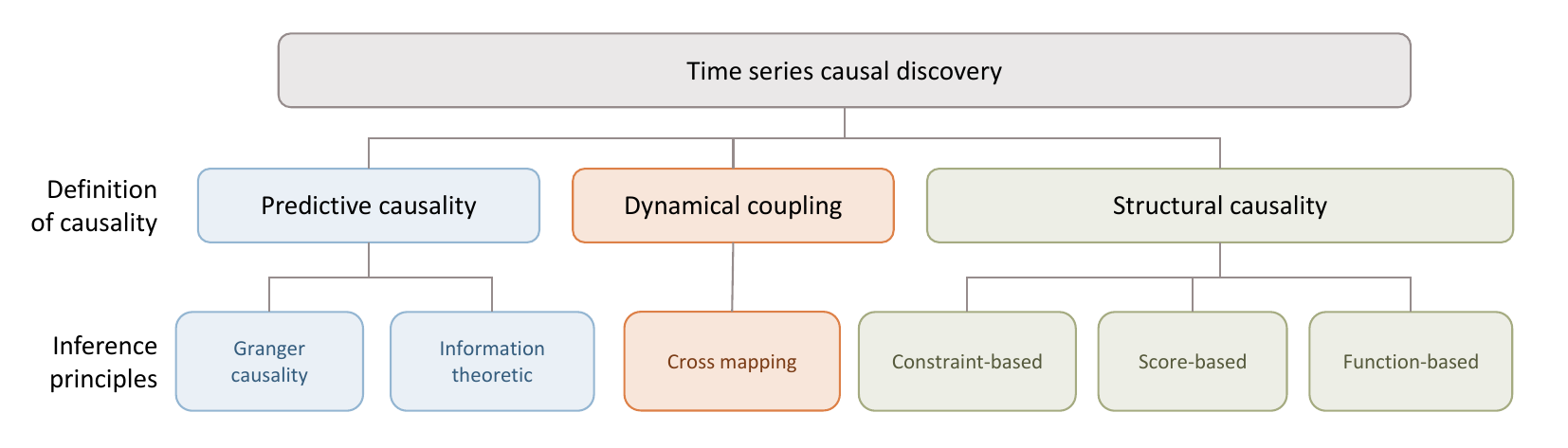}
    \caption{A visual taxonomy of common inference principles for \gls{tscd}. Given a definition of causality, there are often several candidate principles for detecting causation. Compare with Table \ref{tab:methods} to see example methods for each inference principle. This taxonomy does not exhaustively show every principle, and there exist methods which are hybrids of multiple inference principles.}
    \label{fig:taxonomy}
\end{figure}

With the above background, we may now formulate the \gls{tscd} problem. Suppose we are given a set of time series $X_{i,t}$, enumerated by  $i=1,...,d$ and recorded at times $t=1,...,T$. In the window graph \gls{tscd} problem, we use data to infer a graph $\mathcal{G}$ whose nodes are $X_{i,t}$ and directed edges $X_{i,s} \to X_{j,t}$ express causal relationships. At its core, \gls{tscd} is a graph topology inference problem; however, a common source of ambiguity is that a ``causal relationship'' can have several possible definitions (Fig. \ref{fig:taxonomy}). Depending on the method, an edge $X_{i,s} \to X_{j,t}$ could mean:
\begin{itemize}
    \item \textbf{Predictive causality:} $X_{i,s}$ is useful for predicting  $X_{j,t}$ \cite{bressler2011wiener}.
    \item \textbf{Coupled dynamics:} The behaviours of $X_{i,s}$ and $X_{j,t}$ are synchronized \cite{sugihara2012detecting}.
    \item \textbf{Structural causality:} $p(X_{j,t}| \mathrm{do}(X_{i,s}=r))$ is not a constant function of $r$\cite{peters2017elements}.
\end{itemize}
These definitions are all related, and may produce the same graphs in common cases, but not always (we provide an example later in equation \eqref{eq:granger_counterexample}). 

Knowing which definition a method assumes is the first step to choosing the right tool for a given problem. For example, Granger causality is often used in computational neuroscience because it provides sufficient information for neuroscientists to study brain activity \cite{bressler2011wiener}. However, structural notions may be preferred in applications where interventions may influence and affect the generation of data \cite{brodersen2015inferring}.

\subsection{Typical assumptions}
Specific \gls{tscd} methods may make specialized assumptions, but two assumptions are common enough to define beforehand. \gls{tscd} methods exist for when these methods are violated \cite{assaad2022survey}, and form a continually active area of research.

\textbf{Causal sufficiency} states that there are no missing causal variables from our graph. Formally, this asserts that for all distinct variables $X_{i,s}$ and $X_{j,t}$ in our system, there does not exist some unobserved process $Z_{u}$, for some time $u$, which is a cause of both $X_{i,s}$ and $X_{j,t}$.

\textbf{Causal faithfulness} means that the random processes adhere to the causal graph. If $X_{i,s}$ and $X_{j,t}$ are statistically independent conditioned on some  $Z_{u}$, then they must be separated in the corresponding causal graph, conditioned on $Z_{u}$ \cite[p.107]{peters2013causal}. This is the converse of the \textbf{Markov} property, which states that graphical separation implies statistical independence \cite[p.101]{peters2013causal}. 

\textbf{Causal stationarity} is the property that the causal model is invariant to translations in time. Hence, $p(X_{i,t}|Pa(X_{i,t}))$ is time-invariant, and expressions such as $X_{1,t-3} \to X_{2,t}$ hold identically for all $t$.\footnote{This is what allows us to use Window Graphs as faithful descriptors.}
Crucially, causal stationarity is not the same as stationarity of the random process. For example, a random walk $Y_t = Y_{t-1} + \varepsilon_t, \varepsilon_t\sim \mathcal{N}(0,1)$ is causally stationary, but it is not statistically stationary since $\mathrm{var}(Y_t)$ grows with time. 

\section{Methods for Causal Discovery with Time Series} \label{sec:methods}
%introduces the the popular classes of methods for inferring causal graphs with time series, grouped by formalism. Methods such as Granger causality \cite{granger1969investigating}, kernel Granger causality \cite{marinazzo2008kernel}, transfer entropy \cite{schreiber2000measuring}, and their extensions, are sometimes called \textit{predictive causality}, using methods of predictive information and entropy to detect causation between stochastic processes. \textit{Dynamical systems} methods such as convergent cross mapping \cite{sugihara2012detecting} and its extensions \cite{ye2015distinguishing, de2020latent,butler2024tangent}, and SINDy \cite{o2023investigating} differ from predictive causality in that their formalism assumes deterministic system dynamics. Finally, many methods have recently been proposed in the causal ML literature, often grouped into \textit{constraint-based}~\cite{runge2019detecting} that use conditional independence tests to build a set of constraining rules to orient graph edges, \textit{gradient-based}~\cite{pamfil2020dynotears} that search the continuous space of graphs with stochastic gradient descent, \textit{function-based}~\cite{hyvarinen2010estimation,peters2013causal} which make assumptions about the functional form of the data generating process underlying the observed data, and \textit{ordering-based}~\cite{sanchez2025causal} which target finding the topological (causal) ordering of the graph.
We now discuss specific methods for \gls{tscd}. To maintain the overview and motivations for the variety of methods, we discuss principles for inferring causation, and a few exemplary methods. We visualize a conceptual taxonomy of how these principles relate in Fig. \ref{fig:taxonomy}, and we present a comparison of methods and their assumptions in Table \ref{tab:methods}.

\begin{table}[ht]\centering
    \small
    \begin{tabular}{lllll}
        \toprule
        \textbf{Principle} & \textbf{Example Method(s)} & \textbf{Graph Type} & \textbf{Dynamics} & \textbf{Causal Sufficiency} \\ 
        \midrule
        Granger causality     & Linear \gls{gc}\cite{granger1969investigating}   & Summary & Stochastic    & Assumed \\
        Granger causality     & Neural \gls{gc}\cite{tank2021neural}   & Summary & Stochastic    & Assumed \\
        Information-theoretic                       & Transfer entropy\cite{schreiber2000measuring}   & Summary & Stochastic    & Assumed \\
        Cross mapping            & CCM\cite{sugihara2012detecting}            & Summary & Deterministic &  Relaxed \\
        Constraint-based & PCMCI\cite{runge2019detecting}          & Window  & Stochastic    & Assumed \\
        Constraint-based & LPCMCI \cite{gerhardus2020high}          & Window  & Stochastic    & Relaxed \\
        Function-based   & VARLiNGAM\cite{hyvarinen2010estimation}       & Window  & Stochastic    & Assumed \\
        Function-based                       & TiMINo\cite{peters2013causal}          & Summary & Stochastic    & Assumed \\
        Score-based    & DYNOTEARS\cite{pamfil2020dynotears}        & Window  & Either        & Assumed \\
        % Topological ordering     & DOTS\cite{sanchez2025causal}           & Window  & Stochastic    & Assumed \\
        \bottomrule
    \end{tabular}
    \caption{Some representative \gls{tscd} methods from each inference principle and a comparison of their key assumptions.}
    \label{tab:methods}
\end{table}

\subsection{Predictive causality}
Granger's approach \cite{granger1969investigating} to \gls{tscd} begins with considering signals as random processes.  The method considers a detection scheme for causation between pairs of signals, and then the entire graph can be inferred from the results of these pairwise tests. Granger's method hinges on the assumption that an event in the past should be useful to predict the effects of the event in the future, which we refer to as the principle of \textbf{predictive causality}.

Let $X_t$ and $Y_t$ be two discrete-time signals that we are interested in.  We say that $X_t$ Granger-causes $Y_t$ if two conditions are satisfied:
\begin{enumerate}
    \item (Bivariate \gls{gc}) Let $X_{\leq t} = \{X_t, X_{t-1}, \cdots\}.$ At time $t$, conditioning on the past of $X_t$ improves the prediction of future values of $Y_t$, in the sense that
    \begin{equation}
        \label{eq:GC}
                \text{Var}( Y_{t+1} | X_{\leq t}, Y_{\leq t}) <
                \text{Var}( Y_{t+1} | Y_{\leq t})
    \end{equation}
    \item (Conditional \gls{gc}) There is information in $X_t$ not found in other signals. For all (possibly multivariate) signals $Z_t$ available in our system, different from $X_t$ and $Y_t$,  the bivariate \gls{gc} inequality persists after conditioning:
    \begin{equation}
                \text{Var}( Y_{t+1} | X_{\leq t}, Y_{\leq t}, Z_{\leq t}) <
                \text{Var}( Y_{t+1} | Y_{\leq t}, Z_{\leq t})
    \end{equation}
\end{enumerate}
Granger originally used this definition with linear autoregressive models to build a detector. In this case, one first selects the window length hyperparameter $Q$ to be a positive integer. We then fit two order-$Q$ autoregressive models, $M_1$ and $M_2$:
\begin{align*}
    &M_1: &Y_{t} &= \sum_{q=1}^Q \alpha_q Y_{t-q} + \varepsilon_t,
    \\
    &M_2: &Y_{t} &= \sum_{q=1}^Q \alpha_q Y_{t-q} +\sum_{q=1}^Q \beta_q X_{t-q} + \varepsilon_t,
\end{align*}
where $\varepsilon_t$ is white Gaussian noise. 
The detector concludes that $X_t$ Granger-causes $Y_t$ if $M_2$ is the best-fitting model, typically by using an $F$-statistic on the coefficients $\beta_q$, or a likelihood ratio test. Since we are producing one detection result for a pair of time series, Granger causality is often considered to learn a summary graph (Fig. \ref{fig:example_dag})

In using \eqref{eq:GC}, we assume that causation requires at least one time step to move from the cause to effect. However, if we allow conditioning on $X_{t+1}$ alongside $X_{\leq t}$, then we can model \textbf{instantaneous causation} of the form $X_{t+1} \to Y_{t+1}$. However, instantaneous causation is often a signature of too-low sampling rate. It is possible to ignore these effects when the recorded time series have sufficient temporal resolution \cite{bressler2011wiener}, but it is not always possible to guarantee that the given sampling is sufficient. We comment further on this phenomenon in the section on hyperparameter selection.

\subsubsection{Extensions of \gls{gc}}
Considering the popularity of \gls{gc} across the applied sciences, it is unsurprising that many extensions exist. Granger himself proposed spectral versions of the detectors based on Fourier analysis \cite{granger1969investigating}. Popular extensions use kernel methods \cite{marinazzo2008kernel} or deep neural networks \cite{tank2021neural} to generalize \gls{gc} to nonlinear systems. Additionally, variants of \gls{gc} exist to handle other assumption violations, such as latent confounding \cite{bressler2011wiener}. A substantial body of work has also focused on sparse and regularized formulations of \gls{gc} to address high-dimensional settings where the number of variables approaches or exceeds the number of observations.

\subsubsection{Transfer entropy}
Notably, the approach of transfer entropy \cite{schreiber2000measuring} replaces predictive variance with entropy. Rather than quantifying how much the prediction error of a target variable is reduced by incorporating the past of another variable, transfer entropy measures the reduction in uncertainty about the target's future state. This \textbf{information-theoretic} formulation of predictive causality allows transfer entropy to capture arbitrary statistical dependencies in a non-parametric setting\footnote{In effect, it generalises linear \gls{gc}}. 
Similar to \gls{gc}, many variants of transfer entropy have been developed to accommodate multivariate, nonlinear, non-stationary, and high-dimensional time series. 

\subsubsection{Limitations of \gls{gc}}
Predictive causality, despite its utility, possesses fundamental limitations.
Later definitions of causality distinguish between the system that generated a random process, and the random process itself. 
\gls{gc} and transfer entropy define and detect causation through the properties of probability distributions, and by proxy, from data.
However, a key insight of modern causality research is that different \glspl{dgp} can produce the same probability distributions \cite[\S 4.1.1]{peters2013causal}.
Thus, predictive causality, which uses only statistical information, is insufficient to infer structural causality without additional assumptions. In the following sections, we show other methods that make varying specialized assumptions. 
Notably, under certain conditions, predictive causality \textit{does} imply structural causality \cite[p.202]{peters2013causal}.

\subsection{Dynamical systems}
\label{sec:ccm}
Granger's causality starts with the assumption that time series are random processes, but this is not a universal modelling assumption.
In some domains, there are empirically known laws that govern a system's behaviour which may be expressed using the language of ordinary differential equations or their discrete-time analogues. 
In such cases, the \gls{dgp} for the system is a deterministic dynamical system.

Deterministic dynamics can give rise to behaviours that betray our intuitions from the stochastic setting.
For example, if $X_t$ causes $Y_t$, but not vice versa, then predictive causality would assert that one cannot predict $X_t$ from $Y_t$.
As a concrete example, consider the following system where $X_t$ causes $Y_t$ unidirectionally:
\begin{align}
    \label{eq:granger_counterexample}
    X_t &= 3.9 X_{t-1} (1 - X_{t-1}), 
    \\
    Y_t &= 3.7 Y_{t-1} (1 - Y_{t-1} - 0.2 X_{t-1}). \notag
\end{align}
After some algebra, $X_t$ may be written as a function of $Y_t$ and $Y_{t-1}$ \cite[Suppl. Mat.]{sugihara2012detecting}, 
\begin{equation}
    \label{eq:explicit_xmap}
X_t = \frac{3.9}{0.2} \left( 1 - Y_{t-1} - \frac{Y_t}{3.7 Y_{t-1}} - \frac{1}{0.2} \left( 1 - Y_{t-1} - \frac{Y_t}{3.7 Y_{t-1}} \right)^2 \right).
\end{equation}
Since we can predict $X_t$ from $Y_{\leq t}$, we would detect that $Y_t$ Granger-causes $X_t$, leading to false positives in the inferred summary graph.

In the study of deterministic dynamical systems, the mapping \eqref{eq:explicit_xmap} exemplifies a more general mathematical phenomenon.
Let $\rho$ be a positive integer (called the \textit{embedding dimension}).
We define a multivariate signal,
\begin{equation}
\label{eq:delay_embedding}
\mathbf{m}_t^Y = [Y_t,Y_{t-\tau},...,Y_{t-(\rho-1)\tau}]^\top,
\end{equation}
using the time lags of the signal $Y_t$.
The Takens' embedding theorem  \cite{sugihara2012detecting} states that if $\rho$ is large enough, then \eqref{eq:delay_embedding} embeds the full state of the dynamical system into $\mathbb{R}^\rho$.\footnote{Several versions of the theorem exist, each with their own caveats. In \cite{sugihara2012detecting}, Takens' theorem applies to a generic family of observation functions, provided the system state $\mathbf{Z}_t$ evolves on a compact manifold of dimension $d$. The theorem states that $\rho \geq 2d+1$ lags suffice to reconstruct the manifold, in the sense that $\mathbf{Z}_t$ and $\mathbf{m}_t^Y$ are related by a diffeomorphism.}
In this example, we see that $[Y_{t},Y_{t-1}]$ acts as a proxy for the full state $[X_t,Y_t]$, since $X_t$ is obtained as a function of $[Y_t,Y_{t-1}]$.
The mapping $[Y_t,Y_{t-1}] \mapsto X_t$ is called a \textbf{cross map}.

When the conditions of Takens' theorem are satisfied, a causal link in one direction, $X_t \rightarrow Y_t$, induces a cross map in the reverse direction.
Thus, cross maps are useful to infer the topology of a summary graph. Cross maps could exist spuriously in short recording periods, but they become increasingly unlikely as more data is received \cite{butler2023causal}. As such, the  \gls{ccm} algorithm detects causation between signals by checking for the existence of a cross map in the reverse direction. The original CCM algorithm \cite{sugihara2012detecting} proceeds in three steps: 
First, one creates the ``delay-embedding" vector $\mathbf{m}_t^Y$
by concatenating $Y_t$ with lagged copies of itself. The hyperparameter $\tau$ in \eqref{eq:delay_embedding} is a positive integer, and is used to optimize the sampling rate for cross map detection.
Second, we map vectors $\mathbf{m}_t^Y$ to points $\hat{X}_t$, simulating the cross-map. Originally this was performed using $k$-nearest neighbour regression , but theoretically other non-linear regression methods are suitable as well.
Third, one checks if $\hat{X}_t$ are good estimates of $X_t$, using the Pearson correlation coefficient. 
% $$
% \mathbf{Y}_t^{(\rho)} = [Y_t,Y_{t-\tau},...,Y_{t-(\rho-1)\tau}]
% $$

\subsubsection{Limitations of cross mapping}
The CCM algorithm covers an important blind spot of Granger causality, but at a price.
Takens' theorem comes with strong conditions, and there are many potential failure modes for this method when these assumptions are not met, although a number of these conditions can be tested for with care \cite{butler2023causal}.
Considerations of particular note are that \gls{ccm} can be sensitive to measurement noise, stochastic dynamics, signal non-stationarity, and the selection of hyperparameters $\rho$ and $\tau$. Here, $\tau$ artificially changes the sampling rate of the signal for embedding, while $\rho$ controls the memory of the model (see the section on hyperparameter selection). 
Additionally, one must be aware of the \textit{strong forcing} effect: if the cause signal is a strong driver of the effect signal, then bidirectional causation will be detected instead of unidirectional causation \cite{sugihara2012detecting}.

%%%%%%%%%%%%%%%%%%%%%%%%%%%%%%%%%%%%%%%%%%%%%%%%%%%%%%%%%%%%%%%%%%%%%%%%%%%%%%%%%%%%%

\subsection{Structural causality}
In contrast to the detection-driven approaches of predictive causality and cross mapping, causal discovery in the machine learning literature is often framed as a problem of graph topology inference or structural equation modelling with a \gls{dag} constraint. 
For \gls{tscd}, this \gls{dag} corresponds to a window graph (Fig. \ref{fig:example_dag}).
The starting point for these approaches is typically Pearl's \gls{sem} framework, where causal graphs are used to explain how the system responds to interventions \cite{pearl2009causality}.
Since observational data is more ubiquitous than interventional data, structural causality methods for \gls{tscd} often define their graph in a Pearlian sense, and then proceed to use certain assumptions or hypotheses to make the causal structure identifiable in practice.

Most strategies for inferring causal \glspl{dag} can be grouped into three categories: constraint-based, function-based and score-based methods. 
\textbf{Constraint-based} methods impose a set of rules, or constraints, to orient undirected edges in graph skeletons often obtained from conditional independence tests.
Furthermore, since causal models are not generally identifiable from observations only,\footnote{
% The famous example is that of a multivariate Gaussian system. If a pair of random variables are jointly Gaussian, then one can specify multiple distinct causal models that generate the same observational distribution.
Without any restrictions, one can always find a causal model that explains a given bivariate probability distribution \cite[pp.44-45]{peters2017elements}. Using symmetry, this leads to at least two causal models given a pair of non-independent random variables.} \textbf{function-based} causal discovery methods impose inductive biases that constrain the search space.
These assumptions may restrict the function class or make assumptions about the noise terms in the causal model.
Finally, \textbf{score-based} methods use a heuristic or principle to score causal models by a given preference, which can be used to resolve certain non-identifiabilities, and then proceed to infer causal models through optimization.
These differences between constraint, function and score based methods apply to static systems, but also carry over to structural causality approaches for systems of time series.

\subsubsection{Constraint-based methods} These approaches treat causal discovery as a statistical elimination process.
Starting from a complete undirected graph, these methods iteratively prune edges whenever a valid \gls{ci} relationship is found. 
In the context of time series, the most prominent representative of this family is the PCMCI (Peter and Clark Momentary Conditional Independence)  framework~\cite{runge2019detecting}. 
PCMCI addresses the specific challenges of time-series data, namely high dimensionality and strong autocorrelation through a specialized two-stage procedure. 
\textit{Stage 1: PC$_1$ Selection}. The algorithm begins with an adapted version of the Peter-Clark (PC) algorithm to identify a preliminary ``superset'' of parents for each variable. To maintain scalability in time-series settings, PCMCI employs (i) critical optimizations, such as temporal filtering where it restricts the search space to ``past-to-present'' relationships.
(ii) correlation prioritization, where it utilizes a greedy heuristic that prioritizes candidate edges with the highest initial correlation. Thus by performing iterative \gls{ci} tests on these prioritized candidates, the algorithm efficiently prunes the search space to identify a draft set of potential causal drivers, $\mathcal{B}(X_{i,t})$, for every variable in the system.
\textit{Stage 2: \gls{mci}}. To mitigate the high false-positive rates and biases typical of autocorrelated signals, the \gls{mci} test is applied. This stage acts as a rigorous  filter where an edge between a lagged variable $X_{i,t-\tau}$ and the target $X_{i,t}$ is retained only if it satisfies the following condition:
\begin{equation}
    X_{i,t-\tau} \indep X_{i,t} |\underbrace{ \mathcal{B}(X_{i,t}) \backslash \{ X_{i,t-\tau} \}}_{\text{Rival Causes}}, \underbrace{\mathcal{B}(X_{i,t-\tau}).}_{\text{History Causes}}
\end{equation}
The power of this formula lies in its two-fold conditioning strategy:
    Accounting for rival causes controlling for all other suspected drivers of the effect, ensuring the link is not redundant or mediated by another contemporary variable.
    Also by conditioning on the parents of the cause itself the algorithm "neutralizes" the variable's own historical trajectory. This specific step prevents the signs of autocorrelation from being misidentified as new causal influences.
The strength of PCMCI lies in its flexibility regarding the underlying data distribution. Depending on the chosen \gls{ci} test (e.g., partial correlation, Kernel-based measures), the framework can capture linear, nonlinear, and even nonparametric relationships. 
Standard PC algorithms often struggle with time series because the conditioning sets grow rapidly with the maximum lag $\tau$. PCMCI’s \gls{mci} stage uses a significantly smaller, more optimized conditioning set, which enhances statistical power and reliability in finite samples.

\textit{Practical considerations}. Significance level $\alpha$ for \gls{ci} tests and max lag $\tau$ are important hyperparameters; PCMCI scales polynomially with the number of variables and $\tau$. 
%provided output is both a causal graph and causal effects.

\subsubsection{Function-based methods}
These methods restrict the function spaces used in inferring causal models to make them identifiable from observational data. 
Under this alternative assumption,
this allows researchers to distinguish between causal directions that might otherwise be statistically indistinguishable based purely on conditional independencies that can be evaluated with observational data (e.g., $X_t \to Y_t$ vs. $Y_t \to X_t$). 
Function-based discovery is based on the principle that, if the causal direction is $X_t \rightarrow Y_t$ for some pair of random processes $X_t$ and $Y_t$, then there exists a functional relationship $Y_t = f(X_t, \varepsilon_{Y,t})$ involving error/noise variables $\varepsilon_{Y,t}$ (hidden variables that cannot be determined by the observed causes). However, constraints on the function space of such equations and the probability space of error terms will imply that there is no $(g, \varepsilon_{X,t})$ such that $X_t = g(Y_t, \varepsilon_{X,t})$ from $g$ in the chosen function space and $\varepsilon_{X,t}$ drawn from the chosen probability space. This asymmetry allows us to infer directionality.

In the context of time-series models, we require methods for handling a group of time series $\{X_{i,t}:i=1,...,d\}$. 
VARLiNGAM ~\cite{hyvarinen2010estimation}, which is a Vector Auto-Regressive extension of a tabular method known as Linear Non-Gaussian Acyclic Model (LiNGAM), is a powerful approach that bridges classical time-series analysis and causal discovery.
It is built upon two foundational functional assumptions: restricting link functions to the linear family, and, most critically, strictly non-Gaussianity of the error terms.
It extends the standard \gls{var} model—which typically only captures lagged dependencies—by incorporating the ability to identify instantaneous effects occurring within the same time step.
Because functions are linear, the observed signals $X_{i,t}$ can be treated as a linear mixture of independent causal influences. 
Under the assumption of non-Gaussianity, one can use \gls{ica} to ``unmix'' these signals into their independent source components (errors), uniquely revealing the true causal ordering of variables.
TiMINo~\cite{peters2013causal} expands function-based discovery into the non-linear domain by focusing on the Additive Noise Model (ANM) family.
This assumes the structure $X_{i,t} = f( Pa(X_{i,t})) + \varepsilon_{X,t}$ in the time-series setup, where the effect is a function of its parents plus an error term.
A causal structure is deemed plausible only if the resulting noise is statistically independent of the suspected causes (verified by \gls{ci} tests on the residuals).
The algorithm is also highly flexible as it can be instantiated using various signal processing tools, including Generalized Additive Models or Gaussian Processes.
%
%Unlike i.i.d. data, where linear Gaussian models are non-identifiable, TiMINo can identify linear Gaussian causal models specifically because of the temporal information available in time series. It identifies directionality by ensuring the noise of the effect does not correlate with the entire temporal history of the true cause.
%
A useful feature of TiMINo is its conservative nature.
If the data does not fit any tested additive noise model or if multiple models are equally plausible, it returns a `do not know' result rather than a false positive.

\textit{Practical considerations}. Function-based methods are ideal when the physical mechanism of the system is partially known (e.g., sensor output is an additive function of its inputs).
However, they are sensitive to model assumptions, that is, if the true system is non-additive on the error terms, the algorithm should not find any valid causal links.

\subsubsection{Score-based methods}
These methods formulate \gls{tscd} by using a score function which, given a candidate causal model, quantifies the model's fit to a dataset and adherence to other criteria. 
A popular family of score-based methods are based on \textbf{differentiable search}, an approach that casts \gls{tscd} as a continuous optimization task. 
This approach was pioneered first for static graphs, and introduced for time series by the DYNOTEARS~\cite{pamfil2020dynotears} framework, which bypasses the NP-hard nature of combinatorial graph search by using differentiable constraints of adjacency and acyclicity.
The core innovation is the treatment of the adjacency matrix as a continuous real-valued matrix rather than a binary one.
The goal is to minimize a reconstruction loss (typically Mean Squared Error) while enforcing a structural constraint that ensures the resulting graph is a \gls{dag}. 
To achieve this without discrete search, DYNOTEARS introduces a smooth, differentiable acyclicity penalty:
\begin{equation}
    h(\textbf{W}) = \text{Tr}(e^{\textbf{W} \circ \textbf{W}})-d=0,
\end{equation}
where $\textbf{W}$ is the weighted adjacency matrix of a graph, $\circ$ denotes the Hadamard product, and $d$ is the number of variables. $h(\textbf{W})=0$ if and only if the graph is a \gls{dag}, allowing the use of standard gradient descent or augmented Lagrangian optimizers.
DYNOTEARS~\cite{pamfil2020dynotears} extends this principle to the temporal domain by modelling time series as a \gls{var} process that accounts for both instantaneous and time-lagged relationships.
The model assumes the signal $X_{i,t}$ is a linear combination of its own past and the current values of other variables:
% \begin{equation}
%     X_t = W^T X_t + A^T X_{t-1} + \dots + A_p^T X_{t-p} + \varepsilon_t,
% \end{equation}
\begin{equation}
    \textbf{X}_t = \textbf{W}^\top \textbf{X}_t + \textbf{A}_1^\top \textbf{X}_{t-1} + \dots + \mathbf{A}_Q^\top \textbf{X}_{t-Q} + \boldsymbol{\varepsilon}_{t},
\end{equation}
where $\mathbf{X}_t = (X_{1,t},...,X_{d,t})^\top$,  $\textbf{W}$ and $\textbf{A}_q,q=1,...,Q$, capture instantaneous and lagged relationships respectively.  The optimization objective becomes a multi-objective loss function:
\begin{equation}
    \min_{\textbf{W}, \textbf{A}_1,\cdots,\mathbf{A}_Q} \mathcal{L}(\textbf{W}, \textbf{A}_1,\cdots,\mathbf{A}_Q) + \lambda_W \|\textbf{W}\|_1 + \sum_{q=1}^Q \lambda_{{A},q} \|\textbf{A}_q\|_1  + \rho h(\textbf{W}),
\end{equation}
where $\mathcal{L}$ is the least-squares loss, $\|\cdot\|_1$ promotes sparsity (similar to LASSO \cite{tibshirani1996regression}), and $h(\textbf{W})$ ensures the instantaneous graph remains acyclic.\footnote{Since the matrices $A_q$ describe past-to-future causation, we only require an acyclicity constraint on $\mathbf{W}$.}

\textit{Practical considerations}. One of the selling points of DYNOTEARS is scalability, as gradient-based search can converge to a solution substantially quicker than combinatorial approaches. DYNOTEARS is notably not invariant to rescaling data, so it may not be suitable for data sets with varying amplitudes or scales across signals. 
% DYNOTEARS is particularly powerful when one suspects intra-sample dependencies that other methods may miss, but also assumes stationarity. 
Careful tuning of the penalty coefficients to avoid overfitting or cycles might be also necessary.

\subsection{Methods for latent confounding}
In the previous sections, we mainly discussed methods that assume causal sufficiency, which is the absence of unobserved processes that affect the causal relationships of interest. Methods based on predictive causality and structural causality are particularly sensitive to confounding. Notably, and perhaps controversially, cross mapping's definition of causality does not formally invoke causal sufficiency, since it is derived from Takens' theorem rather than a structural causal model. In practice, however, a strong unobserved common driver can still induce false positives for \gls{ccm} between otherwise unrelated variables, so the underlying vulnerability persists despite the differing theoretical foundation. In this section, we briefly mention some methods that \textit{do} adjust for confounding, to demonstrate how this is done in practice. 

LPCMCI (Latent PCMCI) \cite{gerhardus2020high} is a constraint-based method that relaxes the assumption of causal sufficiency, explicitly accommodating unobserved latent processes.
Where PCMCI outputs a \gls{dag}, LPCMCI outputs a partial ancestral graph (PAG). 
Compared to \glspl{dag}, PAGs include an extra edge type, $X_{i,s} \leftrightarrow X_{j,t}$, which expresses bidirectional latent confounding. 
LPCMCI extends PCMCI's two-stage procedure with an iterative, order-independent search: candidate parent sets and ancestral relationships are refined across repeated passes, since without causal sufficiency a single top-down pass (as in PCMCI) can no longer be guaranteed to converge on the correct skeleton.

The use of PAGs is intended to relax the causal sufficiency assumption.
% Similar to the time-series Fast Causal Inference (tsFCI) family, 
% which generalizes Spirtes et al.'s original FCI algorithm \cite{spirtes2000causation} to temporal settings, 
% LPCMCI relaxes the causal sufficiency assumption.
SVAR-FCI \cite{entner2010causal} and tsFCI \cite{malinsky2018causal} also output PAGs using conditional independence tests adapted to lagged and, in the case of SVAR-FCI, structural vector-autoregressive assumptions.
%As with PCMCI's relationship to constraint-based discovery generally, 
These methods trade the strong identifiability of the causal-sufficiency assumption for weaker guarantees: an edge $X_{i,s} \leftrightarrow X_{j,t}$ in the output PAG does not identify the confounder itself, only its existence and approximate temporal footprint, so downstream tasks such as effect estimation cannot proceed as directly as they would from a fully identified \gls{dag}.

\textit{Practical considerations.} Confounder-robust methods are markedly more computationally demanding than their \gls{dag}-based counterparts, since the search must track ancestral and latent-confounding possibilities jointly rather than a single edge type; this typically limits their use to systems with modest numbers of variables.
They are the appropriate choice when the analyst has good reason to suspect unmeasured drivers — as is common in observational settings such as the ICU, discussed in Section V-B — and cannot rule this out by design of the study.

% DM, KB (3-4 pages)
\section{Best Practices for Evaluation and Tuning}
\label{sec:Best Practices for Evaluation and Tuning}
%discusses challenges of designing and evaluating time series causal discovery methods when working with real data. 

\subsection{Hyperparameter selection}
%First, we introduce the basics of hyperparameter selection specific for the time series setting, where parameters like sampling rate and window size are common across formalisms \cite{butler2025rethinking}. 

The performance of causal discovery algorithms is rarely `plug-and-play'.
Their behaviour is governed by hyperparameters, that is, the algorithmic options that define the model's sensitivity, complexity, and temporal resolution. 
While some hyperparameters are method-specific (e.g., the architecture of a neural network), others are shared across almost all \gls{tscd} methods. 
This is especially true of those which control the temporal resolution with which we observe the underlying physical system~\cite{butler2025rethinking}.

\textbf{Window length}. The performance of TSCD methods depends on the amount of memory in the model. The amount of memory is typically controlled by a parameter, such as the window length $Q$ (or similarly the embedding dimension $\rho$).
% In signal processing, we often define a window length $Q$. In the context of causal discovery, this is expressed through the minimum lag ($\tau_{min}$) and maximum lag ($\tau_{max}$). 
% Specifically, if $\tau_{min} = 1$, then $Q = \tau_{max}+1$.
% If $\tau_{min} > 1$, we have the option of either using $Q = \tau_{max}+1$, or using a delay filter to process a time-delayed input signal with a smaller $Q$.
% whose $\tau_{min}=1$ with window length is $Q = \tau_{max} - \tau_{min}$. 
Selecting the optimal value for $Q$ is a balancing act between physical realism and statistical power. 
\textit{Low Q}. 
If the window is too short, the model will fail to capture slow-propagating influences. 
For example, a chemical process with high latency might show no causal link if the window expires before the effect reaches the sensor. 
\textit{High Q}. Setting a very high $\tau_{max}$ increases the number of lagged variables the model must consider. 
This leads to two major issues: (A) Memory and processing requirements scale  with the number of lags, (B) Including irrelevant lags adds noise to the conditioning sets, which can obscure true causal signals and increase the false discovery rate~\cite{runge2019detecting}.
Ideally, $Q$ should match the longest physical delay in the system, though this is rarely known a priori.

\textbf{Sampling rate}. 
The sampling rate $f_S$ of the recording process determines many properties of the discrete-time signals that we process with \gls{tscd} methods. In many cases, we begin with a high-resolution recording, and then apply a downsampling factor $k$ before processing the signals:
\begin{equation}
    k = \frac{f_S^{original}}{f_S^{new}}
\end{equation}
While downsampling is a standard technique to reduce dimensionality and suppress high-frequency noise, it can significantly alter causality in a number of ways.
First, modest downsampling can increase detection power by smoothing out stochastic fluctuations that do not contribute to the underlying causal mechanism. 
Second, if $k$ is too high, the propagation delay between cause and effect may become smaller than a single discrete time step. This collapses a clear time-lagged relationship into an instantaneous (zero-lag) link. 
Third, when causal influences are compressed into the same time index due to aggressive downsampling, algorithms may struggle to identify the correct directionality or, worse, misidentify the structure entirely~\cite{butler2025rethinking}, while \cite{hyttinen2016causal} proved that if the system is sampled at lower sampling rate, the resulting data is causally underdetermined.

% \begin{figure}
%     \centering
%     \includegraphics[width=0.65\linewidth]{figs/SPM_varyK.pdf}
%     \caption{ \textcolor{red}{(PLACEHOLDER) An experiment in which we vary the sampling rate for a data set input to a Granger causality detector. (A figure of this sort will be useful to visualize hyperparameters.}}
%     \label{fig:placeholder}
% \end{figure}

\subsection{Evaluation metrics}
%We next introduce the common evaluation metrics for the task, including classification-based scores (F1, precision and recall) and topological scores (structural Hamming distance, structural interventional distance). 

To probe whether a particular discovery method is suitable to your application, we must quantify how well a predicted graph aligns with the ground truth. In signal processing, we are used to metrics like Mean Squared Error (MSE) or Bit Error Rate (BER), but causal discovery requires a shift toward topological and classification-based accuracy. Evaluating a discovered causal graph $\hat{G}$ against a ground-truth graph $G$ involves checking both the presence of edges and their orientation. Table~\ref{tab:metrics} provides a summary of the evaluation metrics discussed in this section.

\textbf{Classification-based scores}. At its simplest, causal discovery can be viewed as a binary classification task: \ie\textit{Does an edge exist between two variables?} 
We use the standard metrics of (P)recision, (R)ecall, and the (F1)-score to evaluate the predicted adjacency matrix.
\begin{align*}
    \text{R} = \frac{TP}{TP + FN}, && \text{P} = \frac{TP}{TP + FP}, && \text{F1} = \frac{2 \times \text{R} \times \text{P}}{\text{R} + \text{P}},
\end{align*}
where TP is True Positives, FN False Negatives and FP False Positives. 
$R$ measures the ability to find all true causal links. $R < 1$ indicates missed connections (FNs). 
$P$ measures the reliability of the detected links. $P < 1$ indicates spurious connections or false alarms (FPs). 
$F1$ is the harmonic mean of the two, providing a single robust metric for balanced evaluation. 
All three scores range from 0 to 1, where 1 represents a perfect reconstruction of the graph's skeleton and orientation.

\textbf{Topological scores}. While classification scores treat every edge independently, topological scores account for the global structure of the graph. The \gls{shd} is the most widely used metric in the literature\cite{tsamardinos2006max}. Let $G = (V, E)$ denote the ground-truth causal graph and $\hat{G} = (V, \hat{E})$ denote the estimated graph over the vertex set $V$. The \gls{shd} measures the minimum number of edge insertions/deletions required to transform $\hat{G}$ into $G$:
\begin{equation}
\label{eq:shd}
\text{SHD}(\hat{G}, G) = \underbrace{|E \setminus \hat{E}|}_{\text{Missing}} 
+ \underbrace{|\hat{E} \setminus E|}_{\text{Extra}} 
%+ \underbrace{|\{(i, j) \mid (i \to j) \in E \land (j \to i) \in \hat{E}\}|}_{\text{Reversed}}
\end{equation}
When working with directed graphs, a reversed edge poses ambiguity for implementing SHD. 
The formula in \eqref{eq:shd} penalizes reversed edges twice: the correct edge is missing and also the reversed edge is extra. Alternatively, SHD  can be defined as the minimum number of insertions, deletions \textit{and edge reversals} required to transform $\hat{G}$ into $G$, which yields a different formula from \eqref{eq:shd}.
Undirected edges ($i - j$), which can be returned by methods outputting Markov equivalence classes, will be marked as needing to be reversed and hence contribute to a penalty of 1.

A significant limitation of SHD is that it is location-agnostic as it treats all mistakes equally. However, for practitioners interested in causal effects (e.g., If I perturb signal $X$, what happens to $Y$?), some mistakes are much costlier than others. The \gls{sid} addresses this by focusing on interventional distributions \cite{peters2015structural}. It counts the number of pairs $(i, j)$ where the predicted interventional distribution $P_{\hat{G}}(y_j \mid \text{do}(x_i))$ differs from the true one $P_{G}(y_j \mid \text{do}(x_i))$.
\begin{equation}
\label{eq:SID}
    \text{SID}(G, \hat{G}) = \sum_{i=1}^{p} \sum_{j \neq i} \mathbb{I} \left( P_{\hat{G}}(y_j | \text{do}(x_i)) \neq P_{G}(y_j | \text{do}(x_i)) \right),
\end{equation}
where $\mathbb{I}$ is the indicator function. Unlike \gls{shd}, \gls{sid} is asymmetric ($SID(G, \hat{G}) \neq SID(\hat{G}, G)$). A low \gls{sid} suggests that the graph is interventionally reliable, even if the \gls{shd} is not zero. Rather than direct evaluation of \eqref{eq:SID}, SID is computed via a graphical adjustment criterion in practice \cite{peters2015structural}.

\begin{table}[ht]
    \centering
    \begin{tabular}{llll}
    \toprule
    \textbf{Metric} & \textbf{Range} & \textbf{Target} & \textbf{Intuition} \\
    \midrule
    Precision & $[0,1]$ & $\uparrow$ & Avoiding false causal links \\
    Recall & $[0,1]$ & $\uparrow$ & Detecting existing causal links \\
    F1 & $[0,1]$ & $\uparrow$ & Recall-Precision balance \\
    SHD & $[0, p^2]$ & $\downarrow$ & Number of structural errors \\
    SID & $[0, p(p-1)]$ & $\downarrow$ & Number of causal effect errors \\
    \bottomrule
    \end{tabular}
    \caption{Summary of evaluation metrics for determining the accuracy of an inferred causal graph.  $p$ is the number of nodes in the graph. For a window graph with $Q$ past lags and $d$ time series, $p=(Q+1)d$.}
    \label{tab:metrics}
\end{table}

\subsection{No free lunch}
%We discuss how the pursuit of a single best method is a mirage, and how the ranking of methods is contingent on the dataset \cite{assaad2022survey}, leading to the conclusion that a method’s performance often reflects how well its inherent assumptions align with the dataset's specific dynamics \cite{runge2019inferring, montagna2023assumption, butler2023causal}. 

% A central challenge in \gls{tscd} is that inferring structure from strictly observational data is a fundamentally ill-posed problem. Multiple different \gls{dag}s can often represent the exact same observational distribution. These are known as Markov Equivalence Classes (MECs), which are sets of graphs that are statistically indistinguishable without further information. To break the symmetry of an equivalence class and identify a single true graph, we must impose assumptions on the \gls{dgp}. However, identifiability is never free. Every assumption that makes a model easier to solve also restricts the types of data it can accurately describe. For instance, if one assumes linear relationships and non-Gaussian noise (as in VARLiNGAM), one gains the ability to identify instantaneous links. However, if the actual signal is nonlinear or Gaussian, the algorithm may produce confident but entirely incorrect results. Furthermore, because different methods are bespoke to specific sets of assumptions, there is no single algorithm that outperforms all others across all datasets~\cite{assaad2022survey}. A method that excels in linear financial econometrics may fail miserably in nonlinear physiological signal monitoring.
When we introduced constraint and score-based methods, we noted that the inference of causal models without constraints is ill-posed \cite[pp.44-45]{peters2017elements}. Like much of statistics, the inference of causation will always require making assumptions. Assumptions such as linearity or stationarity can be tested for, to some degree. However, \gls{tscd} methods are often justified by more specialized assumptions. For example, CCM assumes that the observed signal dynamics can be explained by a compact attracting manifold \cite{sugihara2012detecting}, which is not directly testable from noisy data \cite{butler2023causal}. 
Even worse, foundational assumptions such as causal sufficiency (which posits that there are no unmeasured confounding variables) are entirely untestable and the validity of these assumptions rests heavily on the judgement of human experts. Without careful consideration of these assumptions in a given application, \gls{tscd} becomes unjustified.

% A defining characteristic of causal discovery, and one that distinguishes it from standard black-box machine learning, is that its core assumptions are generally untestable. One cannot simply test if the data is faithful or if there are no latent confounders. This necessitates a human-in-the-loop approach~\cite{runge2019inferring}, where domain experts (e.g., mechanical engineers or clinicians) must work alongside method developers to determine which assumptions (e.g., stationarity, linearity) are physically plausible for the system being studied.

% While we cannot prove that assumptions hold, some methods are more robust than others in the event when some of their assumptions are violated~\cite{montagna2023assumption}. Alternatively, one can look for ``stability''. That is, if a causal link disappears when we slightly change the sampling rate, shift the time window, or add a small amount of noise, that link is likely a statistical artifact rather than a structural reality. Robust methods are those that demonstrate consistency across different subsets or perturbations of the signal.

\subsection{All shades of benchmarking}
%This leads to a discussion on the merits and pitfalls of benchmarking, contrasting the "perfect" but often biased ground truth of synthetic data with the realistic complexity of real-world data (which usually lacks a ground truth) \cite{reisach2021beware, weichwald2020causal, causalrivers2025}.

Because \gls{tscd} methods are inherently assumption-driven, their performance is inextricably linked to the datasets used for evaluation. A method $M$ that excels on one benchmark may fail on another simply because the underlying \gls{dgp} of the latter violates $M$'s core assumptions. Consequently, the goal of benchmarking is not to find a universal winner, but to rigorously map the applicability boundaries of different algorithms. To achieve this, practitioners typically categorize evaluation data into three distinct types (see Table~\ref{tab:benchmarks} for a summary).

\textbf{Simulations}. Fully synthetic data allows for total control over the \gls{dgp}. Researchers can precisely change parameters like noise levels, non-linearity, and coupling strength to test how a method handles specific assumption violations. The clear advantage of this approach is high internal validity and infinite ground truth availability. However, synthetic data is often criticized for lacking the messiness of real-world signals. Furthermore, recent research has shown that some simulators can be exploited, where an algorithm appears to perform well by picking up on unintended statistical artifacts (e.g., marginal variances) rather than the actual causal structure~\cite{reisach2021beware}.

\textbf{Semi-synthetic}. To bridge the gap between simulation and reality, semi-synthetic datasets have become increasingly used (e.g., CausalTime~\cite{cheng2024causaltime}). These are typically generated in one of two ways. (A) Using domain-specific equations (e.g., climate models, power grid dynamics, or biological pathways) to generate data that follows known physical laws. (B) Using generative models to learn the distribution of real-world data, then enforcing structural constraints to ensure a verifiable ground truth exists for evaluation.

\textbf{Real-world}. The gold standard for testing real-world impact is data collected from physical systems. However, obtaining a definitive causal ground truth for complex systems (like global weather or the human brain) is notoriously difficult. While quality real-world benchmarks have historically been sparse, new community-driven efforts, such as the CausalRivers dataset~\cite{causalrivers2025}, are providing much needed high-fidelity signals with verified causal links.

\textit{Practical considerations}. In the end, all types of benchmarks discussed above have a role in \gls{tscd} development and assessment. Simulations can be used to investigate the method’s limits and sensitivity to hyperparameters. Semi-synthetic approaches can demonstrate performance in a realistic but controlled environment. Whereas real-world data can prove practical applicability and robustness. The best practice is to include at least one semi-synthetic or real-world case study to demonstrate impact, supported by well-controlled synthetic experiments to diagnose algorithmic failures.

\begin{table}[ht]
    \centering
    \begin{tabular}{lll}
    \toprule
    \textbf{Benchmark Type} & \textbf{Primary Purpose} & \textbf{Recommended Usage} \\
    \midrule
    Simulations & Diagnostic & Investigate the method’s limits, sensitivity to hyperparameters, \\ & &and behaviour relative to a known ground truth \\
    Semi-synthetic & Complexity & Demonstrate performance in a realistic but controlled environment \\
    Real-world & Impact & Prove practical applicability and robustness \\
    \bottomrule
    \end{tabular}
    \caption{Summary of benchmark types, their primary purposes, and recommended usage.}
    \label{tab:benchmarks}
\end{table}

\section{Impact and Application Across Domains}
\label{sec:impact}

%%%%%%%%%%%%%%%%%%%%%%%%%%%%%%%%%%%%%%%%%%%%%%%%%%%%%%%%%
% Zhihua for section beginning %
%%%%%%%%%%%%%%%%%%%%%%%%%%%%%%%%%%%%%%%%%%%%%%%%%%%%%%%%%
% The preceding sections have established the formal foundations of \gls{tscd}, from the graphical representations that encode temporal causal structure (Section~\ref{sec:preliminaries}), the algorithmic principles that estimate these graphs from observational data (Section~\ref{sec:methods}), to the evaluation practices and inherent trade-offs that govern their reliable use (Section~\ref{sec:Best Practices for Evaluation and Tuning}). 
The previous sections establish the theory and best practices of \gls{tscd}, but we said very little about what the causal graphs produced by \gls{tscd} are actually useful for. As we observe with \gls{gc}, prediction alone cannot reveal which variables are actual drivers of system behaviour rather than correlates. TSCD closes this gap, turning complex records into interpretable knowledge for forecasting, intervention planning, and model diagnosis. Its value is not restricted to any one field, but lies in a common framework for reasoning about mechanism when experiments are limited and robustness matters as much as accuracy.
Which family from Section \ref{sec:methods} is preferred in each domain is not arbitrary: it tracks whether interventions are feasible, whether confounding is a live concern, and how many variables must be handled at once, corresponding to the axes introduced in Table \ref{tab:methods}.

\begin{figure}[ht]
    \centering
    \includegraphics[width=0.9\linewidth]{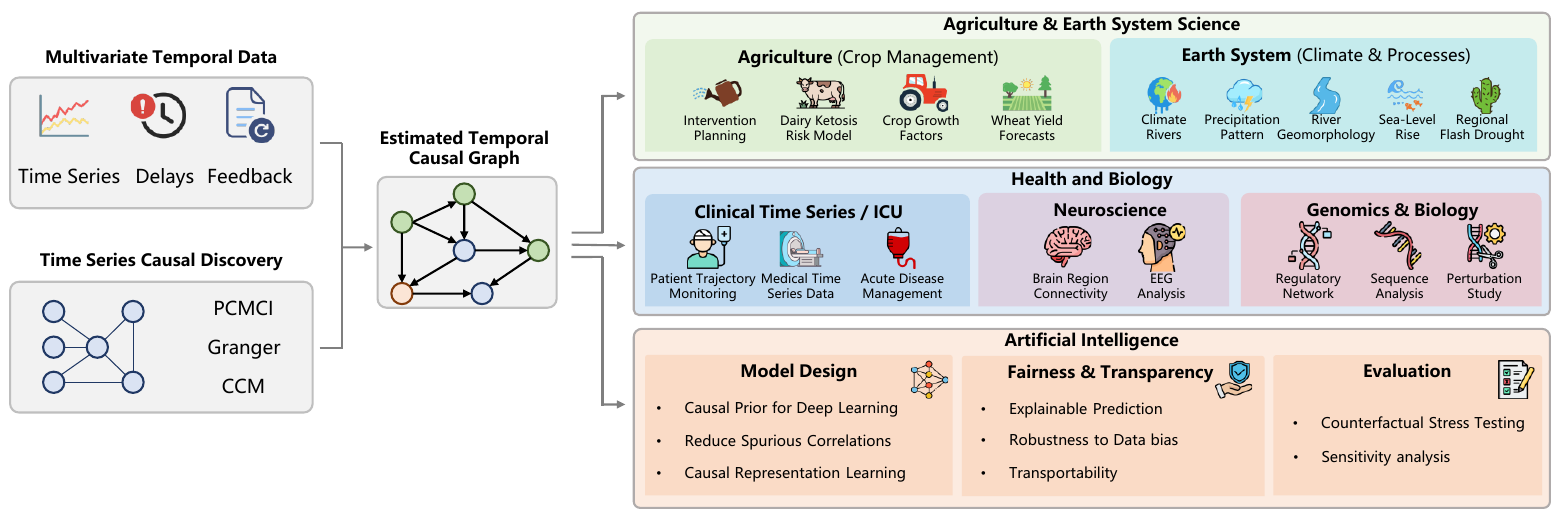}
    \caption{This diagram demonstrates the representative application domains of \gls{tscd}.}
    \label{fig:key applications}
\end{figure}

In this section, we demonstrate the transformative impact of \gls{tscd} across three broad application domains, as illustrated in Fig.~\ref{fig:key applications}. 
% We begin with agriculture and Earth system sciences, where the temporal, continuous nature of environmental data makes temporal causal analysis essential for disentangling interacting physical processes. The second case study is in healthcare, where longitudinal patient data, from intensive-care monitors, EEG signal to genetics, demands \gls{tscd} that goes beyond correlation to identify actionable clinical pathways. Finally, we examine how the outputs of causal discovery are increasingly being leveraged by AI systems as structural priors, improving the safety, robustness, and interpretability of downstream machine learning models. In each application scenario, practitioners face multivariate time series generated by complex, partially observed systems. Standard predictive models can exploit statistical regularities in such data, but they cannot distinguish genuine causal mechanisms from spurious associations. Causal discovery fills precisely this gap, which estimates the underlying causal graph when the true structure is unknown, transforming opaque temporal data into structured, interpretable knowledge that can guide interventions.

%%%%%%%%%%%%%%%%%%%%%%%%%%%%%%%%%%%%%%%%%%%%%%%%%%%%%%%%%
% Feng's Section for Agriculture and Earth System: %
%%%%%%%%%%%%%%%%%%%%%%%%%%%%%%%%%%%%%%%%%%%%%%%%%%%%%%%%%
\subsection{Agriculture and Earth system science}
In agriculture and Earth system sciences, it is common to encounter time series generated by systems with delayed interactions, feedback loops, and seasonalities. 

In agriculture, \cite{ushio2023ecological} combine intensive rice-plot monitoring with environmental DNA and cross-mapping-style analysis to identify organisms influencing rice growth, arguing for cross mapping's suitability for deterministic, low-noise dynamical attractors, rather than the stochastic regime \gls{gc} assumes. \cite{lehmann2020potential} take a constraint-based approach instead, following \cite{runge2019inferring}, to extract precursors of Moroccan wheat yields months ahead of harvest. This is a natural fit given that controlled interventions on continental-scale climate drivers are impossible, and constraint-based methods make no parametric assumption about the underlying dynamics. They additionally restrict the search using physical climate knowledge as a structural prior, illustrating how domain knowledge can narrow an otherwise underdetermined search.

In Earth system science, where replicated interventional experiments are often infeasible at scale, Runge \emph{et al.} argue that TSCD methods are crucial tools for deducing causal insights from otherwise purely observational data \cite{runge2019inferring}. In particular, constraint-based methods like PCMCI are now standard in this field. \cite{yang2023finding} applies it to long-term U.S. precipitation records to identify gateway regions governing precipitation variability, while \cite{kumar2024causal} use causal effect networks built from nearly four decades of data to identify global drivers of regional flash droughts, achieving predictive ability up to four weeks in advance.
% Sun \emph{et al.} further map global soil-moisture--precipitation coupling and propose process-based metrics for model evaluation \cite{sun2025causal}.

These studies in agriculture and Earth system science show that \gls{tscd} is valuable not only because it improves prediction, but also because it permits causal analysis where no interventions are possible.

\subsection{Health science and biology}
These fields represent perhaps the most consequential application domains for \gls{tscd}. 
Many processes in biology are dynamic, evolving gradually over time. Time series in healthcare are also common, as hospitals record patient vital signals or more general electronic health records regularly. Questions about how to best treat a patient are causal in nature, and so \gls{tscd} is a valuable tool for answering such questions.

Consider that an ICU patient generates high-frequency, confounded time series with causal structures that may shift as a disease progresses. Constraint-based TSCD methods are useful for the recovery of causal antecedents of hypertension and hypoxia \cite{srikishan2023causal}. This is precisely the setting where causal sufficiency is least defensible, motivating the need for confounder-robust variants over standard PCMCI. Similarly, the REACT framework applies Granger-style causation to predict acute kidney injury 16 hours earlier than existing guidance-based pathways \cite{zhong2025causal}. 

In neuroscience, Granger-style causal discovery is often used for processing brain signals \cite{barnett2009granger}. Barrett et al. \cite{barrett2012granger} apply permutation-debiased spectral Granger causality to source-localized cingulate cortex recordings from seven subjects under propofol anaesthesia, finding that bidirectional causal influence increased during loss of consciousness in the $\beta$ and $\gamma$ bands with higher inter-subject consistency than phase synchrony. In contrast, GANGO (Greedy Adjacencies and Non-Gaussian Orientations) \cite{rawls2023computational} uses a different approach by decomposing EEG time--frequency surfaces into oscillatory and aperiodic components, then a custom constraint-based causal discovery algorithm to the time-frequency partitions. At the social scale, \cite{silfwerbrand2024directed} applies PCMCI  to hyperscanning EEG from three leader--follower pairs, recovering 21 inter-brain leader-to-follower directed paths and 13 in the reverse direction.

In genomics, gene regulatory network inference pushes TSCD to its limit: tens of thousands of variables observed at only a handful of irregular timepoints. BETS applies conditional \gls{gc} via a sparsity-regularized vector autoregressive model, recovering a network of nearly 2,800 genes and over 30,000 edges from lung cells exposed to a glucocorticoid over 12 hours~\cite{lu2021causal}. In this case, \gls{gc}'s polynomial rather than combinatorial scaling is what makes inference tractable.

% Gene regulatory network (GRN) inference is among the most demanding instantiations of temporal causal discovery. Variables in GRN are tens of thousands of gene transcripts and the directed edges encode which gene causally drives which other. The time series span only a handful of irregularly sampled points, encoding knowledge essential for understanding immune regulation, drug response, and disease mechanisms~\cite{marku2023time}. For bulk RNA-seq, BETS (Bootstrap Elastic net regression from Time Series)~\cite{lu2021causal} applies conditional Granger causality via a vector autoregressive model regularized by an elastic net penalty. Applied to A549 lung cells exposed to the glucocorticoid dexamethasone over 12 hours, BETS recovers a directed network of 2,768 nodes and 31,945 edges, with transcription factors strongly enriched among causal genes. 

By learning directed causal graphs from temporal observations, these applications transform multi-modal patient and biological trajectories into structured, interpretable knowledge. This grounds treatment planning, patient stratification, and precision medicine in mechanisms rather than correlations alone.

%%%%%%%%%%%%%%%%%%%%%%%%%%%%%%%%%%%%%%%%%%%%%%%%%%%%%%%%%
% Feng's Section for AI: %
%%%%%%%%%%%%%%%%%%%%%%%%%%%%%%%%%%%%%%%%%%%%%%%%%%%%%%%%%
\subsection{Artificial Intelligence}
Beyond domain-specific applications, TSCD is increasingly being used to build better AI systems. Instead of simply outputting a causal graph, this field considers how they can be operationalized as design constraints, determining what variables are and are not reliable for decision making. As a concrete example, TSCD methods based on the PCMCI framework \cite{runge2019detecting} were used to identify causally-relevant inputs for a neural prediction model \cite{iglesias2024causally}. This points to a broader role for TSCD in building AI systems that are robust under intervention.

Across all three domains, the choice of method family is dictated less by domain identity than by the availability of interventions, the plausibility of causal sufficiency, and dimensionality, reflecting the recurring decision axes of Table \ref{tab:methods}.

\section{Research Outlook}
\label{sec:outlook}
Despite continual progress, technical challenges to \gls{tscd} remain. Typical issues are present, such as scalability, robustness to non-stationarity, and computational efficiency. There are also classical problems in causality that apply to \gls{tscd}, such as unobserved confounding. One open area of research lies in \textit{anticipatory systems}, where outcomes appear to precede their causes (often due to biological mechanisms with internal ``clocks''). Additionally, there is still a gap between the tooling and empirical performance, with most \gls{tscd} methods underperforming on benchmarks like CausalRivers (see Table 3 in \cite{causalrivers2025}). It is still unknown whether these issues can be resolved with better methods, or if there are fundamental limits on what TSCD can achieve. The underlying reasons can only be understood by analysing the physical and temporal realities that generated these time series. The signal processing community has a wealth of experience dealing with these realities, and unique insights into how they can be overcome.

Recent work has proposed new paradigms for TSCD as well. Topological ordering methods have been proposed as a scalable tool for TSCD, as they divide the graph inference problem into topological ordering search and spurious edge pruning, allowing them to handle larger graphs~\cite{sanchez2025causal}. Amortized causal discovery improves efficiency by training an encoder-decoder to map observed time series directly to graph structures in a single forward pass, bypassing expensive per-dataset optimization. However, these models remain bound to the distribution of systems or shared dynamics seen during training~\cite{lowe2022amortized}. Foundation models extend beyond single-domain amortization by pre-training large architectures, often solely on synthetic graph distributions, to learn a domain-agnostic mapping from raw temporal data to causal DAGs~\cite{stein2024embracing}. While amortized methods speed up inference within a familiar system class, foundation models target zero-shot transfer across entirely unseen domains, scaling performance with model size and pre-training diversity toward general-purpose causal discovery.

\section{Conclusion}
\label{sec:conclusion}
In this tutorial, we have introduced the \gls{tscd} problem, and discussed a number of exemplary methods  for how to solve it. 
Many of these methods rely on various modelling assumptions and philosophies, and we have attempted to treat this material in a unified manner to make the connections and conflicts between these methods more clear. Moreover, we have commented on the challenges of this endeavor, including the trade-offs made in selecting methods and hyperparameter tuning. 
In general, there is still no general-purpose \gls{tscd} method, so in practice we must be careful to select the right approach for our data. Finding general-purpose pipelines for \gls{tscd}, that can automatically analyse a given problem context, design and test methods to be used, and deploy those solutions, remains an open problem. Progress towards this problem-solution mapping requires understanding the strengths, limitations and assumptions of each \gls{tscd} method, and our work is moving in this direction.

Downstream from the theory, solving these problems also has real impacts. Causal reasoning allows intelligent systems to make intelligent decisions and to interact with their environments in a robust and efficient manner. Humans often use causal reasoning implicitly, but for machines to do so they will need infer causation from data. In this sense, \gls{tscd} plays a crucial role in the future deployment of autonomous systems like robots. We already discussed the growing relevance of \gls{tscd} to applications in agriculture, Earth science, neuroscience, and healthcare in the previous section. For technology, \gls{tscd} is also playing a crucial role in the development of new AI systems.

As signal processing evolves from monitoring complex systems to actively manipulating them, unraveling time-dependent causality is no longer just a theoretical pursuit; it is a prerequisite for safe and reliable artificial intelligence.

---------------------------------------------------
% Acknowledgments
% ---------------------------------------------------
%
\section*{Acknowledgment}
This work was supported by the UKRI AI programme and the Engineering and Physical Sciences Research Council, for CHAI - Causality in Healthcare AI Hub [grant number EP/Y028856/1].

---------------------------------------------------
% References
% ---------------------------------------------------
%

\bibliographystyle{IEEEtran}
\bibliography{refs}

@inproceedings{cheng2024causaltime,
title={CausalTime: Realistically Generated Time-series for Benchmarking of Causal Discovery},
author={Yuxiao Cheng and Ziqian Wang and Tingxiong Xiao and Qin Zhong and Jinli Suo and Kunlun He},
booktitle={The Twelfth International Conference on Learning Representations},
year={2024}
}

@article{iglesias2024causally,
  title={Causally-informed deep learning to improve climate models and projections},
  author={Iglesias-Suarez, Fernando and Gentine, Pierre and Solino-Fernandez, Breixo and Beucler, Tom and Pritchard, Michael and Runge, Jakob and Eyring, Veronika},
  journal={Journal of Geophysical Research: Atmospheres},
  volume={129},
  number={4},
  pages={e2023JD039202},
  year={2024},
  publisher={Wiley Online Library}
}

@article{ushio2023ecological,
  title={An ecological network approach for detecting and validating influential organisms for rice growth},
  author={Ushio, Masayuki and Saito, Hiroki and Tojo, Motoaki and Nagano, Atsushi J},
  journal={Elife},
  volume={12},
  pages={RP87202},
  year={2023},
  publisher={eLife Sciences Publications, Ltd}
}

@article{kumar2024causal,
  title={Causal discovery analysis reveals global sources of predictability for regional flash droughts},
  author={Kumar, Sudhanshu and Tian, Di},
  journal={Water Resources Research},
  volume={60},
  number={11},
  pages={e2024WR038391},
  year={2024},
  publisher={Wiley Online Library}
}

@article{yang2023finding,
  title={Finding causal gateways of precipitation over the contiguous United States},
  author={Yang, Xueli and Wang, Zhi-Hua and Wang, Chenghao and Lai, Ying-Cheng},
  journal={Geophysical Research Letters},
  volume={50},
  number={4},
  pages={e2022GL101942},
  year={2023},
  publisher={Wiley Online Library}
}

@article{runge2019inferring,
  title={Inferring causation from time series in Earth system sciences},
  author={Runge, Jakob and Bathiany, Sebastian and Bollt, Erik and Camps-Valls, Gustau and Coumou, Dim and Deyle, Ethan and Glymour, Clark and Kretschmer, Marlene and Mahecha, Miguel D and Mu{\~n}oz-Mar{\'\i}, Jordi and others},
  journal={Nature Communications},
  volume={10},
  number={1},
  pages={2553},
  year={2019},
  publisher={Nature Publishing Group UK London}
}

@article{lehmann2020potential,
  title={Potential for early forecast of Moroccan wheat yields based on climatic drivers},
  author={Lehmann, Jascha and Kretschmer, Marlene and Schauberger, Bernhard and Wechsung, Frank},
  journal={Geophysical Research Letters},
  volume={47},
  number={12},
  pages={e2020GL087516},
  year={2020},
  publisher={Wiley Online Library}
}

@book{peters2017elements,
  title={Elements of Causal Inference: Foundations and Learning Algorithms},
  author={Peters, Jonas and Janzing, Dominik and Sch{\"o}lkopf, Bernhard},
  year={2017},
  publisher={The MIT Press}
}

@article{marinazzo2008kernel,
  title={Kernel method for nonlinear {Granger} causality},
  author={Marinazzo, Daniele and Pellicoro, Mario and Stramaglia, Sebastiano},
  journal={Physical Review Letters},
  volume={100},
  number={14},
  pages={144103},
  year={2008},
  publisher={APS}
}

@article{zhong2025causal,
  title={Causal deep learning for real-time detection of cardiac surgery-associated acute kidney injury: derivation and validation in seven time-series cohorts},
  author={Zhong, Qin and Cheng, Yuxiao and Li, Zongren and Wang, Dongjin and Rao, Chongyou and Jiang, Yi and Li, Lianglong and Wang, Ziqian and Liu, Pan and Che, Hebin and others},
  journal={The Lancet Digital Health},
  volume={7},
  number={9},
  year={2025},
  publisher={Elsevier}
}

@inproceedings{hyttinen2016causal,
  title={Causal discovery from subsampled time series data by constraint optimization},
  author={Hyttinen, Antti and Plis, Sergey and J{\"a}rvisalo, Matti and Eberhardt, Frederick and Danks, David},
  booktitle={Conference on Probabilistic Graphical Models},
  pages={216--227},
  year={2016},
  organization={PMLR}
}

@article{granger1969investigating,
  title={Investigating causal relations by econometric models and cross-spectral methods},
  author={Granger, Clive WJ},
  journal={Econometrica: Journal of the Econometric Society},
  pages={424--438},
  year={1969},
  publisher={JSTOR}
}

@article{schreiber2000measuring,
  title={Measuring information transfer},
  author={Schreiber, Thomas},
  journal={Physical Review Letters},
  volume={85},
  number={2},
  pages={461},
  year={2000},
  publisher={APS}
}

@article{runge2019detecting,
  title={Detecting and quantifying causal associations in large nonlinear time series datasets},
  author={Runge, Jakob and Nowack, Peer and Kretschmer, Marlene and Flaxman, Seth and Sejdinovic, Dino},
  journal={Science Advances},
  volume={5},
  number={11},
  pages={eaau4996},
  year={2019},
  publisher={American Association for the Advancement of Science}
}

@inproceedings{pamfil2020dynotears,
  title={Dynotears: Structure learning from time-series data},
  author={Pamfil, Roxana and Sriwattanaworachai, Nisara and Desai, Shaan and Pilgerstorfer, Philip and Georgatzis, Konstantinos and Beaumont, Paul and Aragam, Bryon},
  booktitle={International Conference on Artificial Intelligence and Statistics},
  pages={1595--1605},
  year={2020},
  organization={Pmlr}
}

@inproceedings{butler2025rethinking,
  title={Rethinking Chronological Causal Discovery with Signal Processing},
  author={Butler, Kurt and Machlanski, Damian and Dimitrakopoulos, Panos and Tsaftaris, Sotirios A},
  booktitle={2025 59th Asilomar Conference on Signals, Systems, and Computers},
  year={2025},
  organization={Institute of Electrical and Electronics Engineers}
}

@article{butler2023causal,
  title={On causal discovery with convergent cross mapping},
  author={Butler, Kurt and Feng, Guanchao and Djuri{\'c}, Petar M},
  journal={IEEE Transactions on Signal Processing},
  volume={71},
  pages={2595--2607},
  year={2023},
  publisher={IEEE}
}

@article{hyvarinen2010estimation,
  title={Estimation of a structural vector autoregression model using non-{Gaussianity}.},
  author={Hyv{\"a}rinen, Aapo and Zhang, Kun and Shimizu, Shohei and Hoyer, Patrik O},
  journal={Journal of Machine Learning Research},
  volume={11},
  number={5},
  year={2010}
}

@article{peters2013causal,
  title={Causal inference on time series using restricted structural equation models},
  author={Peters, Jonas and Janzing, Dominik and Sch{\"o}lkopf, Bernhard},
  journal={Advances in Neural Information Processing Systems},
  volume={26},
  year={2013}
}

@article{sanchez2025causal,
  title={Causal Ordering for Structure Learning from Time Series},
  author={Sanchez, Pedro and Machlanski, Damian and McDonagh, Steven and Tsaftaris, Sotirios A},
  journal={Transactions on Machine Learning Research},
year={2025}
}

@book{pearl2009causality,
  title={Causality},
  author={Pearl, Judea},
  year={2009},
  publisher={Cambridge University Press}
}

@article{assaad2022survey,
  title={Survey and evaluation of causal discovery methods for time series},
  author={Assaad, Charles K and Devijver, Emilie and Gaussier, Eric},
  journal={Journal of Artificial Intelligence Research},
  volume={73},
  pages={767--819},
  year={2022}
}

@article{sugihara2012detecting,
  title={Detecting causality in complex ecosystems},
  author={Sugihara, George and May, Robert and Ye, Hao and Hsieh, Chih-hao and Deyle, Ethan and Fogarty, Michael and Munch, Stephan},
  journal={Science},
  volume={338},
  number={6106},
  pages={496--500},
  year={2012},
  publisher={American Association for the Advancement of Science}
}

@article{reisach2021beware,
  title={Beware of the simulated {DAG}! causal discovery benchmarks may be easy to game},
  author={Reisach, Alexander and Seiler, Christof and Weichwald, Sebastian},
  journal={Advances in Neural Information Processing Systems},
  volume={34},
  pages={27772--27784},
  year={2021}
}

@inproceedings{causalrivers2025,
  title={{CausalRivers} - Scaling up benchmarking of causal discovery for real-world time-series},
  author={Gideon Stein and Maha Shadaydeh and Jan Blunk and Niklas Penzel and Joachim Denzler},
  booktitle={The Thirteenth International Conference on Learning Representations},
  year={2025}
}

@article{bressler2011wiener,
  title={{Wiener}--{Granger} causality: a well established methodology},
  author={Bressler, Steven L and Seth, Anil K},
  journal={Neuroimage},
  volume={58},
  number={2},
  pages={323--329},
  year={2011},
  publisher={Elsevier}
}

@article{srikishan2023causal,
  title={Causal Discovery with Stage Variables for Health Time Series},
  author={Srikishan, Bharat and Kleinberg, Samantha},
  journal={arXiv preprint arXiv:2305.03662},
  year={2023}
}

@article{barrett2012granger,
  title={Granger causality analysis of steady-state electroencephalographic signals during propofol-induced anaesthesia},
  author={Barrett, Adam B and Murphy, Michael and Bruno, Marie-Aur{\'e}lie and Noirhomme, Quentin and Boly, M{\'e}lanie and Laureys, Steven and Seth, Anil K},
  journal={PloS one},
  volume={7},
  number={1},
  pages={e29072},
  year={2012},
  publisher={Public Library of Science San Francisco, USA}
}

@article{rawls2023computational,
  title={A computational framework for {EEG} causal oscillatory connectivity},
  author={Rawls, Eric and Gilmore, Casey and Kummerfeld, Erich and Lim, Kelvin and Nienow, Tasha},
  journal={Proceedings of Machine Learning Research},
  volume={223},
  pages={40},
  year={2023}
}

@article{silfwerbrand2024directed,
  title={Directed causal effect with {PCMCI} in hyperscanning {EEG} time series},
  author={Silfwerbrand, Lykke and Koike, Yasuharu and Nystr{\"o}m, P{\"a}r and Gingnell, Malin},
  journal={Frontiers in Neuroscience},
  volume={18},
  pages={1305918},
  year={2024},
  publisher={Frontiers Media SA}
}

@article{lu2021causal,
  title={Causal network inference from gene transcriptional time-series response to glucocorticoids},
  author={Lu, Jonathan and Dumitrascu, Bianca and McDowell, Ian C and Jo, Brian and Barrera, Alejandro and Hong, Linda K and Leichter, Sarah M and Reddy, Timothy E and Engelhardt, Barbara E},
  journal={PLoS Computational Biology},
  volume={17},
  number={1},
  pages={e1008223},
  year={2021},
  publisher={Public Library of Science San Francisco, CA USA}
}

@inproceedings{malinsky2018causal,
  title={Causal structure learning from multivariate time series in settings with unmeasured confounding},
  author={Malinsky, Daniel and Spirtes, Peter},
  booktitle={Proceedings of 2018 ACM SIGKDD workshop on causal discovery},
  pages={23--47},
  year={2018},
  organization={PMLR}
}

@article{peters2015structural,
  title={Structural intervention distance for evaluating causal graphs},
  author={Peters, Jonas and B{\"u}hlmann, Peter},
  journal={Neural Computation},
  volume={27},
  number={3},
  pages={771--799},
  year={2015},
  publisher={MIT Press}
}

@article{tibshirani1996regression,
  title={Regression shrinkage and selection via the lasso},
  author={Tibshirani, Robert},
  journal={Journal of the Royal Statistical Society Series B: Statistical Methodology},
  volume={58},
  number={1},
  pages={267--288},
  year={1996},
  publisher={Oxford University Press}
}

@article{brodersen2015inferring,
  title={Inferring causal impact using {Bayesian} structural time-series models},
  author={Brodersen, Kay H and Gallusser, Fabian and Koehler, Jim and Remy, Nicolas and Scott, Steven L},
  journal={The Annals of Applied Statistics},
  pages={247--274},
  year={2015},
  publisher={JSTOR}
}

@article{tsamardinos2006max,
  title={The max-min hill-climbing {Bayesian} network structure learning algorithm},
  author={Tsamardinos, Ioannis and Brown, Laura E and Aliferis, Constantin F},
  journal={Machine Learning},
  volume={65},
  number={1},
  pages={31--78},
  year={2006},
  publisher={Springer}
}

@article{barnett2009granger,
  title={Granger causality and transfer entropy are equivalent for {Gaussian} variables},
  author={Barnett, Lionel and Barrett, Adam B and Seth, Anil K},
  journal={Physical Review Letters},
  volume={103},
  number={23},
  pages={238701},
  year={2009},
  publisher={APS}
}

@article{tank2021neural,
  title={Neural {Granger} causality},
  author={Tank, Alex and Covert, Ian and Foti, Nicholas and Shojaie, Ali and Fox, Emily B},
  journal={IEEE Transactions on Pattern Analysis and Machine Intelligence},
  volume={44},
  number={8},
  pages={4267--4279},
  year={2021},
  publisher={IEEE}
}

@article{gerhardus2020high,
  title={High-recall causal discovery for autocorrelated time series with latent confounders},
  author={Gerhardus, Andreas and Runge, Jakob},
  journal={Advances in Neural Information Processing Systems},
  volume={33},
  pages={12615--12625},
  year={2020}
}

@article{entner2010causal,
  title={On causal discovery from time series data using {FCI}},
  author={Entner, Doris and Hoyer, Patrik O},
  journal={Probabilistic Graphical Models},
  volume={16},
  year={2010}
}

@inproceedings{lowe2022amortized,
  title={Amortized causal discovery: Learning to infer causal graphs from time-series data},
  author={L{\"o}we, Sindy and Madras, David and Zemel, Richard and Welling, Max},
  booktitle={Conference on Causal Learning and Reasoning},
  pages={509--525},
  year={2022},
  organization={PMLR}
}

@article{stein2024embracing,
  title={Embracing the black box: Heading towards foundation models for causal discovery from time series data},
  author={Stein, Gideon and Shadaydeh, Maha and Denzler, Joachim},
  journal={AAAI Workshop (AI4TS)},
  year={2024}
}

\vfill

\end{document}